\documentclass[11pt]{amsart}
\makeatletter
\def\NAT@sort{\z@}
\def\NAT@cmprs{\z@}
\makeatother
\usepackage{hyperref}

\usepackage{dsfont}
\usepackage[OT1]{fontenc}
\usepackage[latin9]{inputenc}
\usepackage{mathrsfs}
\usepackage{amsmath}
\usepackage{amssymb}
\usepackage{lmodern}
\usepackage[font=scriptsize]{caption}
\usepackage{textcase}
\usepackage{graphicx, pifont, adjustbox, bm, xcolor}
\usepackage{amsfonts}

\usepackage{comment}
\usepackage{mathtools}
\usepackage{float}
\usepackage{slashed}
\usepackage{ragged2e}

\newcommand{\bg}{\mbox{\boldmath{$ g$}}}

\usepackage{amscd}
\usepackage{amsthm}
\newtheorem{theorem}{Theorem}
\theoremstyle{definition}

\newcommand{\II}{{\rm  I\hspace{-.2mm}I}}
\newcommand{\IIo}{\hspace{0.4mm}\mathring{\rm{ I\hspace{-.2mm} I}}{\hspace{.0mm}}}

\newcommand{\IIIo}{{\mathring{{\bf\rm I\hspace{-.2mm} I \hspace{-.2mm} I}}{\hspace{.2mm}}}{}}
\newcommand{\IVo}{{\mathring{{\bf\rm I\hspace{-.2mm} V}}{\hspace{.2mm}}}{}}

\newcommand{\ext}{{\operatorname{d}}}
 
\newcommand{\pdot}{{\boldsymbol{\cdot}}}

\newcommand{\hh}{{\hspace{.3mm}}}

\newcounter{mnotecount}%[section]
\newcommand{\mnotex}[1]%{}
{\protect{\stepcounter{mnotecount}}$^{\mbox{\footnotesize $\bullet$\themnotecount}}$ 
\marginpar{%\color{red}%
\raggedright\tiny\em
$\!\!\!\!\!\!\,\bullet$\themnotecount: #1} }

\newcommand{\scri}{{\mathscr{I}}}

\newcommand{\m}{{\stackrel{\scalebox{.5}{$\tiny \rm as$}} m}{}}
\newcommand{\akappa}{{\stackrel{\scalebox{.5}{$\tiny \rm as$}} \kappa}{}}

\makeatletter
\let\@bibdataout@init\@empty
\makeatother

\begin{document}

\title[Anti-de Sitter charges and maximal Cauchy hypersurfaces]
{Anti-de Sitter charges and maximal Cauchy hypersurfaces}

\author{A. Rod Gover}
\address{Department of Mathematics, The University of Auckland,
Private Bag 92019,  Auckland 1142, New Zealand}
\email{r.gover@auckland.ac.nz}

\author{Jaros\l aw Kopi\'nski}
\address{Center for Theoretical Physics, Polish Academy of Sciences,
Al. Lotnik\'ow 32/46,  02-668 Warsaw, Poland}
\email{jkopinski@cft.edu.pl}

\author{Andrew Waldron}
\address{Center for Quantum Mathematics and Physics (QMAP), Department of Mathematics, University of California, Davis, CA 95616, USA}
\email{wally@math.ucdavis.edu}

\date{\today}

\begin{abstract}
We construct conserved charges of asymptotic symmetries for  negative-cosmological-constant vacuum spacetimes  from the Neumann data  for  the Hamiltonian and momentum constraints on  maximal conformally-compact Cauchy hypersurfaces.
This gives a general and canonical  initial value treatment of charges for vacuum spacetimes with timelike conformal boundary.
\end{abstract}
 
% insert suggested keywords - APS authors don't need to do this
%\keywords{}

\maketitle

\setcounter{section}{0}

\section{Introduction}

An important problem is the study of the vacuum Einstein equations with cosmological constant~$\Lambda$, 
$$
{Ric}^{g_{\rm ph}}=\tfrac 14 Sc^{g_{\rm ph}} g_{\rm ph}=\Lambda\hh 
g_{\rm ph} \, ,
$$
where the physical metric $g_{\rm ph}$ is Lorentzian  and complete on some four-manifold $X_{\rm ph}$. 
Penrose showed that    asymptotic properties of a spacetime $(X_{\rm ph},g_{\rm ph})$
can be studied by  considering the set~$\scri$  of  limiting points of null geodesics. 
A {\it conformal compactification of a spacetime  $(X_{\rm ph},g_{\rm ph})$} is a Lorentzian manifold $(X,g^o)$ and  a diffeomorphism $\phi:X_{\rm ph} \to \mathring X$ 
such that the metric $ g^o:= (\phi^{-1})^* g_{\rm ph}$ on the interior $\mathring X$ extends  as a  metric 
\begin{equation}\label{singe}
g:=s^2 g^o
\end{equation}
on $\mathring X\cup \scri$, 
where $s$ is any defining function for the conformal boundary $\scri$. The pair $( X,g^o)$ is called a {\it compactified spacetime}. Choosing a different defining function $s'=\Omega s$ with $0<\Omega\in C^\infty(X)$ yields a conformally related metric $g'=\Omega^2 g$, and so in turn a  Lorentzian conformal class of metrics $\bm c:=[g]$ on $X$. 
In this article, we focus only on   the case where $(X_{\rm ph},g_{\rm ph})$
admits a conformal compactification with timelike conformal boundary~$(\scri,\bar {\bm c})$ where $\bar{\bm c}$ is the conformal class of Lorentzian metrics induced by $\bm c$; this implies that  $\Lambda<0$.
A standard example is the anti-de Sitter spacetime for which $X$ is a cylinder $\overline{B^3} \times {\mathbb R}$ with timelike conformal boundary $\scri = S^2 \times {\mathbb R}$.

Charges associated to asymptotic symmetries of timelike boundaries are  key tools for  studying spacetimes, and have been extensively investigated; contributions of particular importance for our work include~\cite{iw,adas,wz,wang,chrnag,bfs,chrh,him,chrmt,michel,Cederbaum,chrwut}. Quite generally, these quantities are associated to spacelike surfaces in the conformal boundary.
An everywhere spacelike  Cauchy surface $S\hookrightarrow  (\scri,\bar {\bm c})$ shall be called a {\it Cauchy cut}.
When the timelike conformal boundary $(\scri,\bar {\bm c})$ admits a conformal Killing vector~$ {\bar \xi}$, Ashtekar and Das~\cite{adas} have constructed a corresponding conserved charge 
\begin{equation} \label{acharge}
Q_{\bar \xi}:=-\int_{(S,\, \,  \tilde \bar {\!\!\bm c}) } \IVo^{\scri}(  \hat {\bar m},{\bar \xi})\, .
\end{equation}
In the above  the  vector $\hat{\bar{m}}$ is the future-pointing unit timelike normal to the Cauchy cut $S$ and~$ \ \tilde \bar {\!\!\bm c}$ is the conformal class induced along $S$.
By virtue of the conformal Killing property of ${\bar \xi}$ and the divergence theorem, the conformally invariant current~$\IVo_{ab}^\scri
{\bar \xi}^b$ is covariantly conserved for {\it any} trace-free, divergence-free symmetric cotensor~$\IVo^{\scri}$ of conformal weight $-1$.
Thus  integration  of $ \IVo^{\scri}(  \hat {\bar m},{\bar \xi})$ with respect to the (weight two) volume form induced by~$\tilde {\bar {\bm c}}$ is well-defined and independent of the choice of Cauchy cut $S$.  When $(\scri,\bar{\bm c})$ is the conformal boundary of a negative cosmological constant vacuum spacetime, for computing  the charges
of asymptotic symmetries, the relevant  tensor~$\IVo^{\scri}$  
is termed the fourth fundamental form of the conformal embedding $\scri\hookrightarrow (X,\bm c)$. It has a simple expression in terms of the electric part of the Weyl tensor of $(X,\bm c)$, and  precisely captures the Neumann data for the vacuum Einstein system with  Dirichlet data given by the conformal manifold~$(\scri,\bar{\bm c})$~\cite{and, gra1,gra2,d2n}. 
To study charges in terms of a gravitational  initial value problem, 
we must  express the Ashtekar--Das charge~$Q_{\bar \xi}$ as an invariant of a 
Cauchy hypersurface $M\hookrightarrow (X,\bm c)$
whose boundary is a Cauchy cut~$S$.
 For anti-de Sitter spacetime $M=\overline{B^3}$ and~$S=S^2$.
 
 Cauchy hypersurfaces  of the compactified spacetime $(X,g^o)$
 %, where~$\phi^* g^o= g_{\rm ph}$, 
 are conformally compact. Recall  here that, in general,  a Riemannian manifold $(M,\tilde g^o)$ is said to be {\it conformally compact} when $M$ is compact with (hypersurface) boundary~$S$ such that, for any  defining function $\tilde s$, the corresponding {\it compactified metric}
$
\tilde g = \tilde s^2 \tilde g^o
$
extends to~$S$ as a metric.
Therefore we must construct invariant charges from Riemannian conformally compact structures. The construction developed below
will be canonical and independent of a reference metric.

\smallskip

Both the extrinsic and the intrinsic geometry of Cauchy hypersurfaces are constrained. A Cauchy hypersurface is maximal when the mean curvature of the singular metric $g^o$ obeys
$$H^{g^o}=0\, .$$
The Hamiltonian and momentum constraints are then
 $$
Sc^{\tilde g^o}=
2\Lambda+
|\IIo^{M}|^2_{\tilde g^o}\, ,\qquad
\tilde\nabla^a_{\tilde g^o} \IIo^{M}_{ab}=0\, .
$$
Here $\IIo^{M}$ is the trace-free part of the extrinsic curvature
 of the embedding $\mathring{M}\hookrightarrow 
(\mathring X,g^o)$. It is determined by the conformal embedding~$M\hookrightarrow (X,\bm c)$ alone.
Also, $\tilde g^o$ is the pullback of~$g^o$ to $M$.

We work with a maximal Cauchy hypersurface $M$ with boundary $S$, whenever such a hypersurface exists. If it is unique, it provides a canonical extension of $S$ into the bulk. Uniqueness is not required for the charge formula below, and indeed our result shows that the Ashtekar--Das charge gives an invariant of maximal Cauchy hypersurfaces anchored at $S$. It is known that such a maximal hypersurface $M$ is smoothly embedded only in the case that the surface $S$ is Willmore in $\scri$---see Section~\ref{MCH}. Uniqueness, existence, and regularity of maximal hypersurfaces in asymptotically anti-de Sitter spacetimes are discussed in~\cite{chrgal}. 
 Observe that when $\Lambda = -3$ and the embedding $M\hookrightarrow (X,\bm c)$ is umbilic, the Hamiltonian constraint becomes the three-dimensional {\it singular Yamabe equation} of~\cite{Mazz}.  The latter is the asymptotically hyperbolic analog of the constant scalar curvature Yamabe condition. In spacetime dimension $d=n+1$, the vacuum Einstein equation is $\operatorname{Ric}^{g_{\rm ph}} =\frac{2\Lambda}{d-2}g_{\rm ph}$, where $\Lambda$ denotes the cosmological constant. We normalize the anti-de Sitter radius to one, so that $2\Lambda = -(d-1)(d-2)$. 
 
\smallskip

The Ashtekar--Das charge  $Q_{\bar \xi}$ probes the entire spacetime rather than its boundary structure alone. Indeed, as discussed above, its integrand depends on the  Neumann data for the vacuum Einstein system, which is only  determined non-locally for given  Dirichlet data along $\scri$. 
This spacetime Neumann data determines the Ashtekar--Das charge.
Our aim is to show that this charge is in fact determined by  the Neumann data for the constraint equations along a maximal Cauchy hypersurface~$M$. Our solution assigns charges for such  asymptotically hyperbolic three-manifolds. 
More precisely, on a conformal three-manifold $(M,\tilde {\bm c})$ with boundary $S$, every solution of a canonical conformal analog of the  gravitational constraint problem
 determines
a {\it Noether operator}~$Q_{\bullet}$ mapping  a pair,  consisting of a scalar  and a tangent vector field on $S=\partial M$, to a charge.

\smallskip

Hamiltonian and momentum constraints are underdetermined as equations on~$M$ for the conformally compact metric~$\tilde g^o$ and trace-free extrinsic curvature~$\IIo^{M}$.
The standard approach to reducing the number of degrees of freedom to those corresponding to solutions of an elliptic system of partial differential equations (PDEs) on $M$ is to  prescribe a conformal class of metrics $\tilde{\bm c}$  and some part of the tensor~$ \IIo^{M}$ (for example  using York decomposition \cite{york}).
 One is then interested in the Neumann data for such an elliptic system.
We would like to find an analog of the fourth fundamental form~$\IVo$ that captures this Neumann data independently of any such decomposition.
Given~$\IIo^{M}$, the 
Hamiltonian constraint is a prescribed scalar curvature equation for a conformally compact metric~$\tilde g^o$. In fact, the boundary curvature scalars $\tilde \pi_{k}$
of~\cite{bcs}, and in particular the critical case $\tilde \pi^{\rm ren}_{n+1}$,
were developed precisely to probe the Neumann data for such systems. 
Here the integer $k$ refers to the order of derivatives transverse to the boundary  and  the superscript ``${\rm ren}$'' is used in deference to the {\it r\^ole}   these tensors play  in  the holographic renormalization program of~\cite{HenSk, bvv, hss, bfs, kl}.
Less developed are analogous tensors $p_{n-1}^{\rm ren}$ for the momentum constraint, the study of which we initiate here.
In particular, we shall construct a pair $(h_4^{\rm ren},p_2^{\rm ren})$ where $h^{\rm ren}_4$ is a weight $-3$ scalar density on $S$ and~$p_2^{\rm ren}$ is a weight~$-2$ one-form. These tensors capture the Neumann data for the Hamiltonian and momentum constraints along maximal Cauchy hypersurfaces; see Section~\ref{Constraints} for details. In particular we show that the  {\it constraint-equation Neumann-type data} $(h_4^{\rm ren}, p_2^{\rm ren})$ 
as defined in Equations~\eqref{h4ren} and~\eqref{p2ren} is well defined even when the smooth Cauchy cut $S$ is not a Willmore surface in $\scri$.
Our result for the Ashtekar--Das charge is  the following theorem,  whose proof may be found in 
Section~\ref{proofth1}.
\begin{theorem}\label{mainth}
Let $(X_{\rm ph},g_{\rm ph})$ be a time-oriented, four-dimensional,
vacuum spacetime with cosmological constant $\Lambda=-3$, so that
$\operatorname{Ric}^{g_{\rm ph}}=-3g_{\rm ph}$.
Let $(X,g^o)$ be the corresponding compactified spacetime with
timelike conformal boundary $\scri$ and  $S\hookrightarrow(\scri,\bar{\bm c})$ be a  Cauchy cut. Suppose that $S=\partial M$ where $M\hookrightarrow X$ is  a compact
Cauchy hypersurface  whose interior
$\mathring M\hookrightarrow X$ is spacelike and maximal
with respect to~$g^o$. 
Then the 
constraint-equation Neumann-type data $(h_4^{\rm ren}, p_2^{\rm ren})$ of 
a solution $(\tilde g^o,\IIo^M)$ to the constraints determines the 
Ashtekar--Das charge corresponding to any conformal Killing field~${\bar \xi}$ on~$\scri$ by the formula
$$
Q_{\bar \xi} = 
-\int_{(S,\, \,  \tilde \bar {\!\!\bm c}) } 
\Big(
 \nu h^{\rm ren}_4  + p_2^{\rm ren}\!\big(  \tilde{  {\bar \xi}} \hh \big)
\Big)\, .
$$
Here the pair $(\nu, \tilde{\bar \xi}\hh )$ consists of the normal and tangential  components of the conformal Killing vector 
${\bar \xi}=\nu\hat{\bar m}+\tilde{\bar\xi}$ along $S$, where $\hat{\bar m}$ is the future-pointing unit timelike normal vector to $S$ in $\scri$. 
%while~$\tilde{{\bar \xi}}$ is the projection of ${\bar \xi}$ onto directions tangential to $S$. 
Moreover
\begin{align*}
h_4^{\rm ren}\:\:&:=\tfrac13\big[
\hat{\tilde n}^a\hat{\tilde n}^b\hat{\tilde n}^c\hat{\tilde n}^d
\tilde\nabla_a\tilde\nabla_b\tilde\nabla_c\tilde\nabla_d\tilde s
+H^{S\hookrightarrow M}\big(2\tilde P_{\hat{\tilde n}\hat{\tilde n}}
-|\IIo^M|_{\tilde g}^2\big)+3\hat{\tilde n}^a\hat{\tilde n}^b\hat{\tilde n}^c\tilde\nabla_a\tilde P_{bc}
\big]\big|_S\, ,\\[2mm]
(p_{2}^{\rm ren})_a &:= \tilde{\bar{\top}} \big[ \tfrac{1}{2} \hat{\tilde n}^b \hat{\tilde n}^c \hat{\tilde n} ^d \tilde \nabla_b \tilde \nabla_c \IIo_{da}^{M} + \tfrac{3}{2} H^{S \hookrightarrow M} \tilde{\bar{\nabla}}^b (i^*\IIo^{M}_{ab}) - \tfrac{1}{2} \IIo^{M}_{ab} \tilde{\bar{\nabla}}^b H^{S \hookrightarrow M} \big]\big|_{S}\, ,
\end{align*}
where $\tilde \nabla$ is the Levi--Civita connection of a compactified metric $\tilde g = \tilde s^2 \tilde g^o$ and $\hat{\tilde n}$ is the unit normal of $S\stackrel i \hookrightarrow (M,\tilde g)$. 
\end{theorem}

\smallskip

In the  geometric set-up described above, $(M,\tilde g^o)$ is an asymptotically hyperbolic manifold. The latter are conformally compact manifolds whose metrics $\tilde g^o$ have asymptotically constant negative scalar curvature. There already exist notions of charges for asymptotically hyperbolic manifolds. In the approaches of Wang~\cite{wang} and Chru\'sciel--Herzlich~\cite{chrh} one first fixes a reference asymptotic model, typically hyperbolic space or a more general static Einstein background, and compares $\tilde g^o$ with this reference metric in a preferred asymptotic chart. Charges may then be written as Hamiltonian boundary charges associated with asymptotic Killing fields of the reference geometry. Subsequent developments include~\cite{chrnag,chrmt,cg,michel}. This construction is conceptually different from covariant phase-space constructions of Hamiltonian generators, such as those of Iyer--Wald and Wald--Zoupas~\cite{iw,wz}, which are closely related to Ashtekar--Das (and more generally, holographic) charges; see {\it e.g.}~\cite{chrwut,him}.

The order of regularity of the various geometric objects appearing in our analysis will play an important {\it r\^ole}. Throughout, the term smooth means $C^\infty$. We will also assume that the four-manifold~$X$ is smooth and that both the compactified metric $g$ and defining function $s$ appearing in Equation~\eqref{singe} are smooth, even though $g^o$ is singular as it approaches $\scri$. Also, the  Cauchy cut $S$ will be assumed to be smoothly embedded. Smoothness of $M$ will be analyzed later. In particular our analysis handles the case when $S$ is not a Willmore surface in $\scri$,
which  leads to a loss of smoothness of the embedding of $M\hookrightarrow X$~\cite{grahwit}.
\begin{figure}[h]
\bigskip
\begin{center}
\begin{picture}(100, 100)(75, 110)
\put(-15, 50){\includegraphics[scale=.5,width=10cm,height=6.2cm]{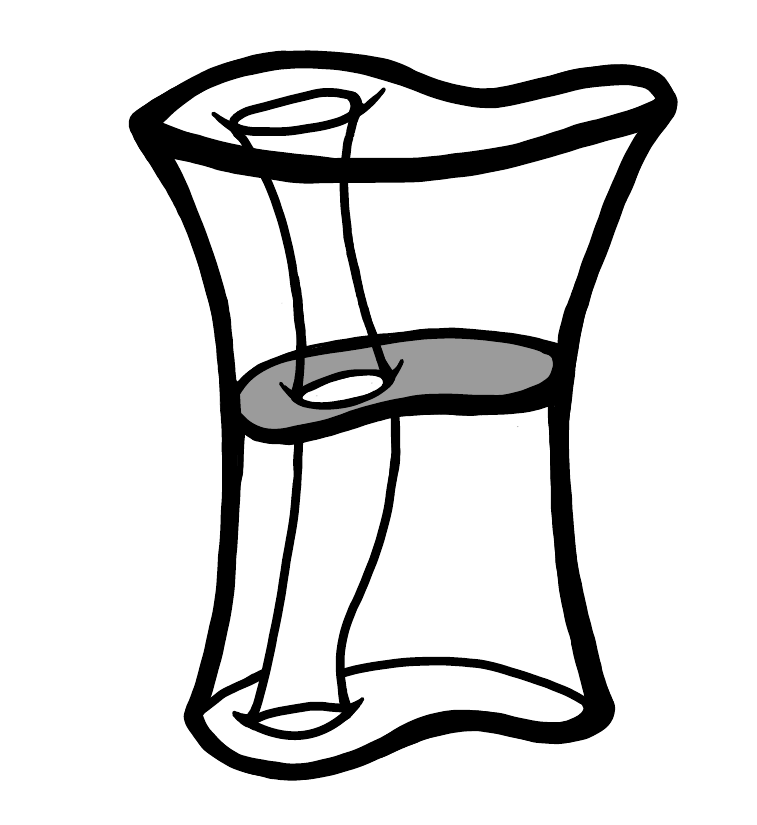}} 
\put(140,105){$(X,g^o)$}
\put(205,105){$(\scri,\bar{\bm c})$}
\put(203,143){\scalebox{.8}{$(S,\tilde{\bar{\bm c}})$}}
\put(139,145.5){\scalebox{.8}{$(M,\tilde g^o)$}}
\put(191.3,153.5){$\big\uparrow$}
\put(189.5,163){$\hat{\bar  m}$}
\put(172.8,143.5){\scalebox{.95}{$\!\hat {\tilde n}\!\!$} \raisebox{.3mm}{$\longleftarrow\, $}}
\end{picture}
\end{center}
\vspace{2cm}
\caption{An artist's impression of a compactified, vacuum, $\Lambda<0$ spacetime $(X,g^o)$ and a maximal Cauchy hypersurface $(M,{\tilde g}^o)$.
The decorations \raisebox{-1mm}{$\tilde{}\,$}, \raisebox{-1mm}{$\bar{}\,$} and \raisebox{-1mm}{$\tilde{\bar{}}\,$} are respectively used to distinguish objects along $M$, $\scri$, and~$S$.}
\end{figure}

\subsection{Conventions}
Here we spell out our notations as well as some key extrinsic and conformal-geometric notions.

\begin{quote}
{\it Non notationes sed notiones.}

\hfill  Carl Friedrich Gau\ss.
\end{quote}

\subsubsection{Tensors}

We will often employ the abstract index notation of Penrose \cite{penr} when handling tensor fields on a smooth $n$-dimensional manifold~$M$. For example $v^a \in \Gamma (TM)$ and $\omega_a \in \Gamma (T^*M)$ are vector and one-form fields. Round and square brackets  denote symmetrization and antisymmetrization, {\it e.g.} $T_{(ab)} \in \Gamma(S^2 T^*M)$ and $T_{[ab]} \in \Gamma(\Lambda^2 T^*M)$. When a metric $g$ is available, we may raise and lower indices, so $T_{ab}=g_{ac}g_{bd}T^{cd}$. 
The trace-free part of a symmetric tensor $T_{ab}$ is denoted by
$\mathring{T}_{ab}:=T_{ab}-\frac{1}{n}g_{ab}T_c{}^c$,
or equivalently by $T_{(ab)^\circ}$. For vectors~$v$ and~$u$ we employ notations $v^2:=g_{ab}v^a v^b=v_a v^a$ and $u.v:=u_a v^a$, and for tensor contractions we denote, for example,~$T_{uv}:=u^a v^b T_{ab}$.
Also, we sometimes employ the standard notations $\operatorname{tr}_g X:=g^{ab} X_{ab}$, $\operatorname{div}_g X:=\nabla^a X_a$, $(\operatorname{div}_g T)_b:=\nabla^aT_{ab}$ and $X^\sharp = g^{-1}(X,\pdot)$. 

The Riemann curvature tensor~$R$
for  a Levi-Civita connection $\nabla^g$ of some metric~$g$ on $M$ 
is  defined~by 
$$[\nabla_a, \nabla_b] v^c =: R_{ab}{}^c{}_d v^d\, .$$
We often drop the superscript $g$ when context makes it clear. The Ricci tensor is the partial trace $Ric_{ab} := R_{ca}{}^{c}{}_{b}$, while the scalar curvature is the full trace, $Sc:= Ric_a{}^a$. In dimensions $n\geq 4$, the Riemann curvature tensor can be decomposed into Weyl and Schouten tensors (the latter defined for $n \geq 3$), $R_{abcd} =: W_{abcd} + 2(g_{c[a}P_{b]d}+g_{d[b}P_{a]c})$, where $Ric_{ab}=: (n-2) P_{ab} + g_{ab} J$ with $J:=P_a{}^a=\frac1{2(n-1)} Sc$. The Cotton tensor is defined as the covariant curl of the Schouten tensor, $C_{abc}:= 2\nabla_{[a} P_{b]c}$.
\subsubsection{Hypersurfaces}
A function $s\in C^\infty(M)$ is said to be {\it defining} for a hypersurface embedding $S\hookrightarrow M$ if $S=s^{-1}(0)$ and~$\ext s$ is nowhere vanishing along $S$. Given a metric $g$ on $M$, let~$\hat{ n} := \frac{(\ext s)^\sharp}{|\ext s|_g}\big|_S$
be the corresponding unit normal vector. The extrinsic curvature,
or second fundamental form, is defined by
$$
\II_{ab} := (\nabla_a-\hat n_a \nabla^g_{\hat n}) \hat n^{\rm ext}_b
|_S=:\nabla^\top_a \hat n_b|_S\, .
$$
Here $\hat n^{\rm ext}$ is any smooth extension of $\hat n$ to a neighborhood of $S$. We also use the notation $\top$ for the action of the (Riemannian signature) projector $\delta^a{}_b - \hat n^a \hat n_b$. The extrinsic curvature of $S$ can be used to relate the intrinsic Riemann curvature tensor $\bar{R}_{abcd}$ of the hypersurface~$S$ to the projection of the Riemann curvature tensor of $M$ via the Gauss equation,
\begin{equation} \label{gauss}
\bar{R}_{abcd} = R^{\top}_{abcd} +\II_{ac} \II_{bd} - \II_{ad}\II_{bc}\, .
\end{equation}
Completely tracing the Gauss equation gives a relation between  scalar curvatures of $S$ and $M$ which, when expressed in terms of traced Schouten tensors, reads
\begin{equation}\label{Ricci}
J - \bar J = P_{\hat{n} \hat{n}} -\tfrac{n-1}{2} H^2 + \tfrac{1}{2 (n-2)} \IIo_{ab} \IIo^{ab}\, .
\end{equation}
Here the extrinsic curvature has been decomposed into its trace-free part $\IIo$ and mean curvature $H:=\tfrac{1}{n-1} \II_a{}^a$.

Contracting the Gauss Equation~\eqref{gauss} with $\bar{g}^{bd}$, then using the scalar Gauss Equation~\eqref{Ricci} yields
\begin{equation}\label{Fialkow}
\IIo_{ac}\IIo^c{}_b-\tfrac{1}{2(n-2)} \bar g_{ab} |\IIo|^2_{\bar g}-\hat n^c \hat n^d W_{cabd}=(n-3)\big(
P^\top_{ab}-\bar P_{ab} + H \IIo_{ab} + \tfrac12 \bar g_{ab} H^2
\big)\, .
\end{equation}

The Codazzi--Mainardi equation relates the covariant curl of the extrinsic curvature of $S$ to the Riemann tensor of $M$,
\begin{equation}\label{CM}
R^{\top}_{abc \hat{n}} = 2 \bar{\nabla}_{[a} \II_{b]c }\, .
\end{equation}
The trace of this  can be used to express the Schouten tensor of $M$ along $S$ in terms of the extrinsic curvature,
\begin{equation}\label{SCM}
P^{\top}_{a \hat{n}} = \tfrac{1}{n-2} \bar{\nabla}_b \IIo^b{}_a - \bar{\nabla}_a H.
\end{equation}
Lastly, we will write $A \stackrel{S}{=} B$  if $A|_S = B|_S$ for some quantities $A$ and $B$.

\subsubsection{Conformal geometry}
A conformal manifold $(M^n,\bm c)$ is a smooth, for simplicity oriented,  
$n$-dimensional manifold with an equivalence class of metrics $\bm c\ni g,g'$ where $g'\sim g$ when
$$
g'=\Omega^2 g\, ,
$$
for some $0<\Omega\in C^\infty(M)$.
The ray bundle of volume forms is denoted ${\mathcal E}M$. Any metric $g\in \bm c$ defines a canonical section $\epsilon^g\in \Gamma({\mathcal E}M)$.
Powers of ${\mathcal E}M$  also make sense, so for any $w\in {\mathbb R}$ we may define the  {\it  weight $w$ conformal density bundle} by
$$
{\mathcal E}M[w]:=({\mathcal E}M)^{-\frac w n}\otimes {\mathbb R}M\, ,$$
where ${\mathbb R}M$ is the trivial line bundle with section space  $C^\infty(M)$. Sections  $\varphi
=f \hh (\epsilon^g)^{-\frac w n} \in \Gamma({\mathcal E}M[w])$ may be labeled by an equivalence class of metric--function pairs $[g,f]=[\Omega^2 g , \Omega^w f]$ because
$
\varphi = (\Omega^w f) (\epsilon^{\Omega^2 g})^{-\frac w n}
$. For any vector bundle $VM$ we denote the weighted bundle $VM[w]:=VM\otimes {\mathcal E}M[w]$. Rather than write cumbersome equations such as $\varphi=[g,f]=0$ we often employ the notation $f\stackrel g = 0$.

An everywhere positive section $\tau \in \Gamma({\mathcal E}M[1])$ is termed a {\it scale}. A scale $\tau$ determines a metric $g_\tau\in \bm c$ such that $\tau=[g_\tau,1]$ and conversely any metric $g$  determines a scale $\tau_g$ . The {\it conformal metric}
is the (tautological) section  
$$
{\bm g}:=\tau^{2} g_\tau=\tau_g^2 g
$$
of $S^2 T^*M[2]$ defined independently of any choice of scale $\tau$ or metric $g$. The Levi--Civita connection~$\nabla^g$ of a metric $g\in \bm c$ also acts on conformal densities $\varphi \in \Gamma({\mathcal E}M[w])$ according to
$$
\nabla^g \varphi := \tau_g^w \nabla (\tau_g^{-w} \varphi)
\, . 
$$
Here $\tau_g^{-w} \varphi \in C^\infty(M)$. 
Note that $\nabla^g \bm g = 0$ and the conformal metric may also be used to raise and lower indices, for example if $\omega_b\in \Omega^1(M)$ then ${\bm g}^{ab} \omega_b \in \Gamma(TM[-2])$.
Finally, if $s$ is any defining function for a hypersurface, %$N\hookrightarrow (M,\bm c)$, 
then the density $\sigma= [g,s]\in \Gamma({\mathcal E}M[1])$ is termed a {\it defining density}.

\section{Smooth extensions}

An efficient way to handle systems of PDEs defined only  on the interior $\mathring M$ of a conformally compact manifold is to search for an equivalent system of equations on the interior that extends smoothly to the compact manifold~$M$. 
%Such a system goes under the   ``almost equation'' moniker. 
To this end, one starts by  describing a conformally compact manifold 
$(M,\tilde g^o)$  by everywhere smooth data 
$(M,\tilde {\bm c}, \tilde \sigma)$, where~$\tilde \sigma$ is a defining density for $S = \partial M$ and,
on the interior $\mathring M$, 
$$
\tilde g^o = \tilde \sigma^{-2} \tilde {\bm g}\, .
$$
This leads to a general method for extending equations to which we attach the adjective {\it almost}.
For our current purposes, given some equation
$E(\tilde g^o)=0$ on the interior $\mathring M$ where $E\in \Gamma(V\mathring M)$, the corresponding almost equation $\mathds{E}=0$ will be defined by 
$$
\mathds{E}
(\tilde{\bm g},\tilde \sigma):= \tilde\sigma^k E(\tilde \sigma^{-2} \tilde{\bm g}) \in \Gamma(VM[k])\, ,
$$
where $k\in {\mathbb{R}}$ is the smallest value for which the right-hand side is well-defined everywhere in $M$.

\smallskip

For example, the {\it almost Einstein equation} obtained from the vacuum Einstein equations
on $\mathring X$
is given by
\begin{equation} \label{aeeq}
 s P+ \nabla n  +  \rho g \stackrel g= 0\, .
\end{equation}
Here $k=1$,  $\sigma=[g,s]$,  $n:=\ext s$, and $\rho :=-\frac14 (\Delta  + J) s$. \\

Another important example for our geometric set-up is the {\it almost maximal equation} for a hypersurface embedding $\mathring M\hookrightarrow (\mathring X,g^o)$,
\begin{equation}\label{AMC}
\left( s  H -  n \left( \hat m \right) \right)|_M \stackrel{g}{=} 0\, ,
\end{equation}
which
is equivalent to $H^{g^o}=0$ along $\mathring M$. The above almost equation also has $k=1$. In  the case that $M$ is a spacelike hypersurface, the vector  $\hat m$ is the future-pointing unit timelike normal of $M\hookrightarrow (X,\bm c)$.

There are also almost analogs of the constraint equations, defined along all of a Cauchy hypersurface $M$ of a conformally compact spacetime $( X,g^o)$. Here the induced conformally compact metric on $M$ is $\tilde g^o=\tilde\sigma^{-2} \tilde{\bm g}$ where the defining density  $\tilde \sigma$ for $S\hookrightarrow M$ is the pullback of $\sigma$ to $M$. Then the {\it almost Hamiltonian} and {\it almost momentum} equations are
\begin{equation} \label{almcon}
|\tilde n|^2_{\tilde g}+2 \tilde \rho \tilde{s} \stackrel{\tilde{g}}{=} 1 -\tfrac16 \tilde s^2
|\IIo^{M}|^2_{\tilde g}\, ,\qquad
\tilde s\tilde\nabla^a \IIo^{M}_{ab}-2\tilde n^a\IIo^{M}_{ab} \stackrel{\tilde{g}}{=}0\, .
\end{equation}
In both cases $k=0$ and $\tilde \rho \stackrel{\tilde{g}}{:=}-\frac13 (\Delta^{\tilde g}  + J^{\tilde g}) \tilde s$. 
Here, $\tilde s$ is the pullback of $s$ to $M$, and $\tilde n = \ext \tilde s$.
We note that when the embedding $M\hookrightarrow (X,{\bm c})$ is umbilic, the almost Hamiltonian equation is equivalent to the three-dimensional {\it almost scalar curvature equation}; see~\cite{Goal} for a general discussion of almost scalar curvature equations. Also observe that  the almost momentum equation is well-defined on all of $M$ even though the trace-free part of the second fundamental form of  $M\hookrightarrow (X,g^o)$  computed with respect to the physical metric is in general singular along~$S=\partial M$.
 
In the case of a Lorentzian conformal four-manifold $(X,\bm c)$ with Cauchy hypersurface $M$ and defining function~$t$, the almost scalar curvature equation becomes\begin{equation}\label{ASC}
m^2 + 2 \kappa t \stackrel g=-1 \, ,
\end{equation}
where $m:=\ext t$  and $\kappa:=-\frac14(\Delta^g + J^g)t$. 
This is the almost equation  (again, with $k=0$) for the dimension four, Lorentzian, singular Yamabe equation.

\section{Fundamental forms}

The Ashtekar--Das charge in Equation~\eqref{acharge} does not appeal to a  reference metric. It is an example of a general method 
%developed in~\cite{CFFs}, 
where instead of writing invariants in terms of (physical, reference)-metric pairs, one instead searches for natural tensors that vanish when evaluated on a class of reference metrics. Recall here that a tensor is said to be natural if it is a map from the space of all Riemannian(/Lorentzian) manifolds taking values in section spaces over the input Riemannian  manifold. The notion of naturality extends to more rococo structures such as the space of conformal hypersurface embeddings.
Fundamental forms of a non-null conformal embedding $\Sigma\hookrightarrow(X,\bm c)$ are a sequence of natural tensors  $\IIo,\IIIo,\IVo,\ldots$, where $\underline{\mathring{\overline{k}}}$ takes values in weighted trace-free symmetric tensors $\Gamma(S^2 T^*\Sigma[3-k])$~\cite{CFFs}. For four-dimensional spacetimes, only the first few fundamental forms are needed.
To describe these  
let  us consider  a  conformal, pseudo-Riemannian four-manifold $(X^4,\bm c)$ as well as an embedded hypersurface
$$
\Sigma_\varepsilon^3\hookrightarrow (X^4,\bm c)\, .
$$
For further concreteness, let us assume that $\Sigma_\varepsilon$ is either everywhere spacelike~($\varepsilon = -1$) or timelike~($\varepsilon=1$). The normalized normal vector $\hat n$ is a vector defined on $\Sigma_\varepsilon$
taking values in weight $-1$ densities, subject to 
$$
\bm g (\hat n,\hat n)= \varepsilon\, .
$$ 
We can now generalize the notation $\top$ to also define the projector onto $\Sigma_{\varepsilon}$, so that
$\top^a{}_b=\delta^a{}_b - \varepsilon \hat{n}^a \hat{n}_b $.

Let us suppose  that $\Sigma_\varepsilon$ is the zero locus of some defining density $\sigma$, in which case we may ask   whether the pseudo-Riemannian metric 
\begin{equation}\label{go}
g^o=\sigma^{-2}\bg\, ,
\end{equation}
defined on $\mathring X:=X\backslash \Sigma_\varepsilon$, solves the vacuum Einstein equations.  
Necessarily then, the scalar curvature must obey
$
\operatorname{sgn}(Sc^{g^o})=-\varepsilon
$.
Furthermore, 
 the  conformal embedding $
\Sigma_\varepsilon^3\hookrightarrow (X^4,\bm c)
$ must now be umbilic, meaning
$$
\IIo^{\Sigma_\varepsilon}_{ab}=0\, .
$$
In addition, along $\Sigma_\varepsilon$, the Weyl tensor of $\bm c$ must also vanish,
$
W_{abcd}|_{\Sigma_\varepsilon}=0
$. This  implies that the tensor 
$$
\IIIo^{\Sigma_\varepsilon}_{ab} := 
W_{\hat n ab\hat n}-\IIo^{\Sigma_\varepsilon}_{c(a}\IIo^{\Sigma_\varepsilon}_{b)^\circ}{}^c
$$
vanishes. The same conclusion holds for higher dimension-$d$ vacuum spacetimes for which the Weyl tensor does not identically vanish along the hypersurface. In fact, the {\it second conformal fundamental form} $\IIo^{\Sigma_\varepsilon}_{ab}$ and 
{\it third conformal fundamental form} $\IIIo^{\Sigma_\varepsilon}_{ab}$ are the first two  of a sequence of $d-2$ conformal fundamental forms that necessarily vanish along~$\Sigma^{d-1}_\varepsilon$ when $g^o$ is Einstein. Indeed, conformal fundamental forms are the~order-by-order local obstructions to the embedding $\Sigma_\varepsilon \hookrightarrow(X,\bm c)$ admitting an (anti-)de Sitter metric whose conformal infinity is the hypersurface $\Sigma^{d-1}_\varepsilon$.

The $d^{\rm th}$ fundamental form in $d$ dimensions is also an invariant of the embedding, but plays a rather different {\it r\^ole} to lower fundamental forms. 
In the case~$\varepsilon = 1$, and $d$ even, it captures global Neumann data 
for the conformally compact vacuum Einstein problem when the
timelike hypersurface~$\Sigma_\varepsilon^{d-1}$ is~$\scri$ ~\cite{d2n}. When $\varepsilon=-1$ and the spacelike hypersurface is Cauchy, it gives the second piece of initial data---the first being 
the induced conformal class of metrics on~$\Sigma_\varepsilon^{d-1}$---required to solve the  vacuum Einstein equations. When those equations hold, the $d^{\rm th}$ conformal fundamental form is both trace- and divergence-free.
Returning to $d=4$ dimensions, 
the fourth conformal fundamental form evaluated on a vacuum solution for either sign of $\varepsilon$ is given by \cite{govk} 
\begin{equation} \label{ivo}
\IVo^{\Sigma_\varepsilon}_{ab} = C_{\hat n ab}^\top \big|_{\Sigma_\varepsilon}\, .
\end{equation}
The above tensor can also be obtained from the leading asymptotics of the electric part of the Weyl tensor~$W_{\hat n ab \hat n}$ \cite{abk}. That this tensor captures the boundary Neumann data for solving the Einstein condition holographically off a timelike boundary underlies the original construction of the Ashtekar--Das charge.

\section{Boundary curvature scalars}\label{BCS}
\noindent

Boundary curvature scalars are the almost scalar curvature equation analogs of conformal fundamental forms. Needed here is the  three-manifold case, so
let $(M^3,\tilde{\bm{c}})$ be a compact conformal three-manifold (of Riemannian signature) with hypersurface boundary 
$$
S\hookrightarrow (M^3,  \tilde{\bm{c}} )\, .
$$
Then by general theory~\cite{ln,AvMcO, ACF,Mazz}, there always exists a unique defining density  
$\sigma_{\rm SY}$
such that, on $M\setminus S$ the metric
$$
\tilde g^o_{\rm SY}={ \sigma_{\rm SY}^{-2} }\ {\tilde {\bm g}}
$$ 
has constant negative scalar curvature $Sc^{\tilde g^o_{\rm SY}}=-6$.
The density $\sigma_{\rm SY}$ is then termed a {\it singular Yamabe scale}. In fact, for hypersurfaces embedded in an $n$-dimensional conformal manifold, the local embedding data alone determines an {asymptotically singular Yamabe} defining density $\sigma_{\rm aSY}$ whose corresponding metric has constant scalar curvature up to terms of order  ${\mathcal O}( \sigma_{\rm aSY}^n)$; see~\cite{ACF, gwln}. Note that $\mathcal{O}(A)$ denotes $A B$ for some smooth quantity $B$.

The question naturally then arises of characterizing when a given defining density $\tilde \eta$ is the singular Yamabe scale. A partial answer to this question is given by a sequence of conformally invariant scalars $\tilde\pi_k$ depending only on the data $S\hookrightarrow (M^3,\tilde{\bm c}, \tilde{\eta})$ \cite{bcs}. Indeed the first of these {\it boundary curvature scalars} is the function
\begin{equation}\label{pi1}
\tilde \pi_1=\tilde\nabla_{\hat{\tilde n}} \tilde{\eta} |_S\, ,
\end{equation}
where $\hat{\tilde n}$ is the unit normal of $S$ in $M$. Higher order examples are density-valued and grow in complexity,
\begin{align}
\tilde \pi_2 &=
( \hat{\tilde n}^a \hat{\tilde n}^b \tilde\nabla_a \tilde\nabla_b \tilde \eta -  H^{S \hookrightarrow M} \tilde\pi_1) \big|_S
\, \in \, \Gamma({\mathcal E} S[-1])\label{pi2}
\, ,\\
\tilde \pi_3|_{\tilde \pi_1=1}&= \left( \hat{\tilde n}^a \hat{\tilde n}^b \hat{\tilde n}^c \tilde\nabla_a \tilde\nabla_b\tilde \nabla_c \tilde \eta +2(\tilde P_{\hat{\tilde n}\hat{\tilde n}} + |\IIo^{S\hookrightarrow M}|^2) \right)\Big|_S
\,  \in\,  \Gamma({\mathcal E} S[-2])
\, .\label{pi3}
\end{align}
These formul\ae\ deserve some explanation. Firstly the scale $\tilde{\eta}$ is only singular Yamabe when $\tilde \pi_1-1=0=\tilde \pi_2$. Were $M$ to have dimension four or higher there would exist a conformally invariant boundary curvature scalar $\tilde \pi_3$. Three dimensions is a special case where one must assume that the metric $\tilde{\bm g}/\tilde{\eta}^2$ is asymptotically hyperbolic, or equivalently $\tilde \pi_1=1$ to write down the ``relative'' conformally invariant third  boundary curvature scalar~$\tilde \pi_3|_{\tilde \pi_1=1}$ given above.
Again, the scale $\tilde{\eta}$ is only singular Yamabe when $
\tilde \pi_3|_{\tilde \pi_1=1}=0$.
In summary, in dimension three, $\tilde\pi_1-1$,~$\tilde\pi_2$ and $\tilde \pi_3|_{\tilde \pi_1=1}$ are obstructions to a defining density being singular Yamabe. A similar picture holds for the first through $n^{\rm th}$ boundary curvature scalars in $n$ dimensions.

Three dimensions is a critical case for the fourth boundary curvature scalar. Even though the PDE obeyed by a scale $\sigma_{\rm SY}$ is second order, the solution  $\sigma_{\rm SY}$ is determined by its zero level set alone. The latter can  be viewed as Dirichlet data which is all that is needed to determine a  solution to this elliptic second-order PDE problem.
The formal expansion does suffer a loss of uniqueness because it is only sensitive to information local to the boundary. The undetermined part of the formal expansion may be viewed as Neumann data, and appears at quartic order in three dimensions.
The unique global solution therefore gives a Dirichlet-to-Neumann map, whose image is captured by the fourth boundary curvature scalar.
To construct that map, in the special case when the conformally compact metric is asymptotically hyperbolic and the embedding $S\hookrightarrow M$ is umbilic ($\IIo^{S\hookrightarrow M}=0$) one uses that  there is an invariant fourth normal order operator, given by 
 \begin{equation*}
\tilde \pi_4|_{\IIo^{S\hookrightarrow M} =0, \tilde{\pi}_1 =1} = \big[ \hat{\tilde n}^a \hat{\tilde n}^b \hat{\tilde n}^c \hat{\tilde n}^d \tilde\nabla_a \tilde\nabla_b \tilde\nabla_c \tilde\nabla_d  \tilde \eta +3 \big( \tilde{\bar{\Delta}}\! -\! \tilde{\bar{J}} \big) \tilde{\pi}_2 + 2 H^{S \hookrightarrow M}(\tilde{\pi}_3 \!+\! \tilde{P}_{\hat{\tilde n}\hat{\tilde n}}) +2 \tilde{P}_{\hat{\tilde n}\hat{\tilde n}} \tilde{\pi}_2
+ 3\hat{\tilde n}^a \hat{\tilde n}^b  \tilde{\nabla}_{\hat{\tilde n}} \tilde{P}_{ab } \big] \big|_{S}\, ,
\end{equation*}
when acting on any {\it smooth} defining density $\tilde \eta$, where $\tilde{\bar J}:=\tfrac12\operatorname{Sc}^{\tilde{\bar g}}$. When specialized to the case that~$\tilde \eta$ is the singular Yamabe scale $\sigma_{\rm SY}$, which implies that $\tilde\pi_1-1$, $\tilde\pi_2$, and $\tilde \pi_3|_{\tilde \pi_1=1}$ vanish, the above formula recovers the Neumann data for the singular Yamabe problem; see~\cite{bcs} for details. 

By construction, in $n$ dimensions, the $(n+1)^{\rm st}$ boundary curvature scalar probes $n+1$ transverse derivatives of the defining density $\tilde \eta$. Therefore, it can also be used to extract Neumann data for more general prescribed scalar curvature problems such as the Hamiltonian constraint on Cauchy hypersurfaces. In particular, this applies when $n=3$ and~$\tilde \eta$ is   the defining density~$\tilde \sigma$  solving the almost Hamiltonian constraint in Equation~\eqref{almcon}, and in that case we will use the special notation
\begin{equation}\label{h4ren}
h_4^{\rm ren}:=\tfrac13 \tilde \pi_4|_{\IIo^{S\hookrightarrow M} =0, \tilde{\pi}_1 =1}\, .
\end{equation}

\section{Maximal Cauchy hypersurfaces}\label{MCH}
We would like to associate a canonical Cauchy hypersurface $M\hookrightarrow X$ to a given Cauchy cut $S\hookrightarrow \scri=\partial X$. For that we impose a maximality condition.
Let us begin with the  embedding~diagram
$$
\begin{array}{ccl}
S&\hookrightarrow &\: M\\[-2mm]
\rotatebox{270}{$\hookrightarrow$}&&\:\rotatebox{270}{$\hookrightarrow$}\\[4mm]
\scri&
\hookrightarrow 
&\!(X,\bm c)\, ,
\end{array}
$$
where $(X,\bm c)$ is a Lorentzian, conformal,  four-manifold with timelike hypersurface boundary~$\scri$ and~$S$ is a Cauchy cut.  The three-manifold  $M$ 
is a Cauchy hypersurface for the compactified spacetime~$(X,g^o)$, where $g^o$ is as described in Equation~\eqref{go}.
 It is natural to ask that $M$ meets $\scri$ orthogonally,
\begin{equation}\label{mn0}
\bg (\hat m,\hat n)|_S=0\, , 
\end{equation}
where the spacelike vector $\hat n$ is the  unit normal to $\scri$ and the timelike vector $\hat m$ is the unit, future-pointing normal to $M$. By orthogonality, the unit normal to \(\scri\hookrightarrow X\) restricted to \(S\) is the unit normal to \(S\hookrightarrow M\); we denote this common normal by \(\hat {\tilde n}\). Likewise, \(\hat{\bar m}=\hat m|_S\). The condition in Equation~\eqref{mn0} naturally  extends  to $M$ 
upon requiring that the embedding
$$
\mathring M\hookrightarrow (\mathring X,g^o)
$$
is maximal; this is evident upon examining the almost maximal Equation~\eqref{AMC}.
To study solutions of the  almost maximal equation it is useful to 
assume that $M$ is the zero locus of some defining density~$\tau=[g,t]$.
Indeed, without losing generality, we may demand that $t$ is  asymptotically singular Yamabe so that the almost scalar curvature Equation~\eqref{ASC} then becomes 
\begin{equation} \label{asymyt}
m^2 + 2 \kappa t \stackrel g=-1 + {\mathcal O}(t^4)\, .
\end{equation}
Here $m:=\ext t$ gives a canonical extension of $\hat m$ to $X$;  again see~\cite{will}. The density $\tau$ is now determined up to order~${\mathcal O}(\tau^5)$. This is the highest order 
to which 
$\tau$ is determined without prescribing further Neumann initial data.
The almost maximal equation for $M$ can  then  be rewritten as
$$
\big(m.n+\kappa s\big) \big|_M \stackrel g= 0\, ,
$$
where again $n:=\ext s$ and $\sigma = [g,s]$ is the defining density for the conformal embedding $\scri\hookrightarrow (X,\bm c)$. The above condition can also be extended as a conformally invariant equation to $X$,
\begin{equation} \label{am2}
m.n + \kappa s + \rho t \stackrel{g}{=} c  t 
%+ b s^3
 \, ,
\end{equation}
where  $\mathcal{C}:=[g,c]$ is some smooth, weight $-1$ density, see~\cite{gwa} for details.

Asymptotics of solutions to the almost maximal equation are well understood~\cite{grahwit, grar, gwa,Alex}. 
In general, the conformal embedding $M\hookrightarrow (X,\bm c)$ may not be smooth, but there always exists a smooth asymptotic solution for the hypersurface $M$ whose defining density $\tau_{\rm as}=[g,t_{\rm as}]$ obeys
\begin{equation}
 \label{am2as}
\m_an^a+\akappa s+\rho t_{\rm as}
\stackrel{g}{=}c\,t_{\rm as}+b s^3\, .
\end{equation}
Here $[g,b]=:\mathcal B$ is a smooth, weight $-3$ density,
$\m_a\stackrel{g}{:=}\nabla_a t_{\rm as}$, and
$\akappa\stackrel{g}{:=}-\tfrac14(\Delta+J)t_{\rm as}$.
  In the case that the interior metric is Einstein, it follows that  $\mathcal{B}|_S$ is proportional to the Willmore invariant~${\mathcal B}^{S\hookrightarrow \scri}$ of the conformal embedding $S\hookrightarrow (\scri,\bar{\bm c})$, that is given by~\cite{grahwit,  gwa}
\begin{equation}\label{Willmore}
{{\mathcal B}}^{S\hookrightarrow \scri}:=
\big(\tilde{\bar\nabla}_a \tilde{\bar\nabla}_b +\varepsilon H^{S\hookrightarrow \scri} \IIo^{S\hookrightarrow \scri}_{ab} + \bar P_{ab}\big)
(\IIo^{S\hookrightarrow \scri})^{ab}\, .
\end{equation}
In the above, $\varepsilon=-1$, but the same formula applies to the Riemannian  embedding $S\hookrightarrow M$ upon setting $\varepsilon = 1$ and replacing $\bar P_{ab}$ by the Schouten tensor of $M$. 
Smoothness of the  embedding $M\hookrightarrow (X,\bm c)$ up to the boundary then requires that the surface embedding $S\hookrightarrow (\scri,\bar{\bm c})$ is Willmore,  meaning that $S$ extremizes the Willmore functional 
$$\int_{(S,\tilde{\bar{\bm c}})}  \big| \IIo^{S\hookrightarrow \scri} \big|^2\, ,$$
 so that its functional  gradient~${{\mathcal B}}^{S\hookrightarrow \scri}$ with respect to compactly supported variations of the embedding vanishes.
 Note that the integrand $ \big| \IIo^{S\hookrightarrow \scri} \big|^2$ is termed the {\it rigidity density} of the conformal embedding $S\hookrightarrow (\scri,\bar{\bm c})$. Indeed, for Riemannian signature embeddings the Willmore functional~\cite{Willmore} is a surface bending energy, while in Lorentzian signature it models  rigid string dynamics~\cite{Polyakov}.
 
Let us list several useful consequences of Equation~\eqref{am2} and its jets along $S$. Firstly, evaluating it on~$S$ implies that the Cauchy hypersurface $M$ meets the conformal boundary orthogonally, as mentioned earlier. After differentiating Equation~\eqref{am2}, and decomposing it into parts normal and tangential to $M$,  we arrive at the following relations,
$$
\mathcal{C} \big|_S = 0   \quad   \mbox{ and }\quad  \hat{\tilde n}^b \IIo^{M }_{a b} \big|_S = 0\, .
$$
After comparing the expressions along $S$ for  two transverse derivatives of~\eqref{am2} with the bulk Laplacian acting on the same equation, one arrives at
$$
 \nabla_{\hat m} c \big|_S \stackrel{g}{=}0 \quad   \mbox{ and }\quad \tilde{\nabla}_{\hat{\tilde n}} H^{M} \stackrel{S}{=}\bar{\nabla}_{\hat{\bar m}} H^{\scri}\, .
$$  
Observe that because $M$ meets $\scri$
orthogonally  and the embedding $\scri\hookrightarrow X$ is umbilic, we have that
\begin{equation} \label{iism}
\II^{S \hookrightarrow M}  = \tilde{\bar{g}}\hh  H^{\scri} \big|_{S}  \  .
\end{equation}
Here $\tilde{\bar{g}}\in \tilde{\bar{\bm c}}$ is a metric on $S$. As a consequence
$$
H^{S \hookrightarrow M}  = H^{\scri} \big|_{S} \:\mbox{ and } \:
\IIo^{S\hookrightarrow M}=0\, .
$$
Similarly, the second fundamental forms are related by surface
pullback. Along $S$ we have
\begin{equation*}
\II^{S\hookrightarrow\scri}=\tilde{\bar{\top}} (\II^M)\big|_S,
\end{equation*}
where $\tilde{\bar{\top}}$ denotes the projector onto directions along $S$. Using also $\hat{\tilde n}^b\IIo^M_{ab}|_S=0$ established above, we obtain
\begin{equation} \label{iism2}
\IIo^{S\hookrightarrow\scri}=
\IIo^M|_S
\:\mbox{ and } \:
H^{S\hookrightarrow\scri}=H^M\big|_S\, .
\end{equation}

\section{Constraint equations on maximal Cauchy hypersurfaces}\label{Constraints}

Our aim is to express the  Ashtekar--Das charge  in terms of the constraint-equation Neumann-type data~$(h_4^{\rm ren}, p_2^{\rm ren})$ of a maximal Cauchy hypersurface $M$. The system of almost Hamiltonian and momentum Equations~\eqref{almcon} can be formulated as  the following boundary problem,  
\begin{equation}\label{BP}
\left\{
\begin{array}{cc}
|\tilde n|^2_{\tilde g}+2 \tilde \rho \tilde{s} \stackrel{\tilde g}{=} 1 -\frac16 \tilde s^2
|\IIo^{M}|^2_{\tilde g}\, ,
&
\qquad\tilde s|_S=0\, ,
\\[3mm]
\tilde s\tilde\nabla^a \IIo^{M}_{ab}-2\tilde n^a\IIo^{M}_{ab} \stackrel{\tilde g}{=}0
\, ,&
\qquad
\IIo_{ab}^M \big|_S=\IIo_{ab}^{S\hookrightarrow \scri}
\, ,
\end{array}
\right.
\end{equation}
where the second boundary condition follows from conditions in~\eqref{iism2}. This coupled system of PDEs is underdetermined, as a simple degree of freedom count quickly verifies: a choice of conformal class~$\tilde{\bm c}$ in dimension three modulo diffeomorphisms gives two field theoretic degrees of freedom. The trace-free symmetric tensor~$\IIo^M$ and defining density~$\tilde \sigma=[\tilde g,\tilde s]$ subjected to the four PDEs above give two further field theoretic degrees of freedom. This yields two canonical pairs corresponding to the physical degrees of freedom of a gravitational field.

Even though the boundary problem  in Equation~\eqref{BP} is underdetermined, there are still good notions of Neumann data.
One approach is to prescribe suitable conformal seed data and use the
Lichnerowicz--Choquet-Bruhat--York conformal method. In the maximal case,
the momentum constraint gives a linear elliptic equation for a vector
potential, independent of the conformal factor, and the Hamiltonian
constraint gives the scalar Lichnerowicz equation for the conformal
factor, see e.g.~\cite{khurikopinski}. Our strategy is to instead
construct invariants of solutions~$(\tilde {\bm c}, \tilde \sigma,\IIo^M)$ to the constraints that probe terms in their asymptotics that are non-local. For the physical spacetime $X_{\rm ph}$, the fourth fundamental form $\IVo$ gives both the non-local Neumann data and the data required to construct charges. We seek  invariants of    a global solution to Equations~\eqref{BP} that not only probe non-local Neumann data, but  also allow us to determine   $\hat {\bar m}^a\IVo_{ab}$, and therefore charges when symmetries are present, along the Cauchy cut.

\smallskip

Any spacetime must give a solution for  the  almost constraint equations. We can therefore probe a spacetime by examining the asymptotics of the corresponding  defining density $\tilde \sigma$ of $S\hookrightarrow (M,\tilde{\bm c})$ and the  conformal weight~1 tensor $\IIo^M$ solving the system in Equation~\eqref{BP}. 
The case ${\mathcal B}^{S\hookrightarrow\scri}=0$, where the embedding $S\hookrightarrow(\scri,\bar{\bm c})$ is Willmore flat,
 is optimal since the embedding $M\hookrightarrow (X,\bm c)$ is then smooth.
 In this section we shall assume that this holds and take up the case where $M$ loses regularity in Section~\ref{notreg}.

\smallskip

We will begin by analyzing the asymptotics of the defining density $\tilde \sigma$ encoded by the Hamiltonian constraint. The boundary curvature scalars described in Section~\ref{BCS} capture the behavior of $\tilde \sigma$ near $S$. The critical boundary curvature scalar $\tilde{\pi}_4$ is only well defined when the defining density~$\tilde\sigma$ is smooth to sufficiently high order.
However, since  $\tilde \sigma$ is the  pullback  of the defining density~$\sigma$ to the smooth hypersurface embedding $M\hookrightarrow(X,\bm c)$, smoothness to all orders is guaranteed, so $\tilde{\pi}_4$ is well defined in the current setting.
\smallskip

Because the physical metric $g_{\rm ph}$ satisfies vacuum Einstein field equations, as discussed earlier, the defining density~$\sigma$ for~$\scri\hookrightarrow X$ satisfies the almost Einstein Equation~\eqref{aeeq}. This in turn implies that the boundary curvature scalars for the defining density~$\tilde\sigma$ induced along $M$ obey
\begin{equation} \label{pi14}
\tilde{\pi}_1 -1= 0 = \tilde{\pi}_2, \quad \tilde{\pi}_3 = - \tfrac{1}{2} \big| \IIo^M \big|_{\tilde{g}}^2 \big|_{S}, \quad h^{\rm ren}_4 =  C_{\hat{\tilde n} \hat{\bar m} \hat{\bar m}} \big|_{S}\,.
\end{equation}
A proof of these relations is presented  in Appendix~\ref{apenh4}. The first shows that $\tilde \sigma$ is asymptotically singular Yamabe to second order. 
In general~$\tilde{\pi}_3$, given by the rigidity density of $M\hookrightarrow(X, \bm c)$, need not vanish; this reflects the fact that the Hamiltonian constraint is not a constant scalar curvature equation, but rather a prescribed scalar curvature equation for the singular metric~$\tilde g^o$.  Recall that the fourth boundary curvature scalar is critical in three dimensions (see the discussion in Section~\ref{BCS}). In fact, by applying the asymptotic analysis of~\cite{gwln,will}, one sees
that  $h^{\rm ren}_4=\frac{1}{3}\tilde{\pi}_4$ captures the Neumann-type data of the Hamiltonian constraint. The above display relates it to the bulk Cotton tensor, and thus in turn to the fourth fundamental form $\IVo$ of $\scri\hookrightarrow (X,\bm c)$.
\medskip

We now consider the formal asymptotics of the trace-free extrinsic curvature $\IIo^M_{ab}$ of $M\hookrightarrow (X,\bm c)$ when it solves the almost momentum constraint. 
Because $M$ is smoothly embedded,  we may make the decomposition
$$
\IIo_{ab}^M\stackrel{\tilde g}{=}\IIo^{M, \textrm{as}}_{ab}+ \mathcal{O} (\tilde{s}^{\ell+1})
$$
where
\begin{equation} \label{komasym}
\IIo^{M, \textrm{as}}_{ab}
:\stackrel{\tilde g}=
\mathring{K}_{ab} + \tilde{s} \mathring{K}^{(1)}_{ab} + \tilde{s}^2 \mathring{K}^{(2)}_{ab} +  \cdots + \tilde{s}^\ell \mathring{K}^{(\ell)}_{ab}  \, .
\end{equation}
Here $\ell \in \mathbb{N}$ and  $\mathring{K}_{ab}$ is {\it any} fixed initial choice of  smooth extension of $\IIo^{S\hookrightarrow \scri}_{ab}$ to  $M$.
Also  $\mathring{K}^{(i)}_{ab}$ (with $i \in\{1,\ldots, \ell\}$) are representatives of smooth  tensor-valued densities of weight $1-i$.
%, which we would like to determine as far as is possible. 
To characterize the asymptotics of~$\IIo^M_{ab}$, we try to obtain that~$\IIo^{M, \textrm{as}}_{ab}$ satisfies 
$$
\tilde s\hh\tilde\nabla^a \IIo^{M, \textrm{as}}_{ab}-2\tilde n^a\IIo^{M, \textrm{as}}_{ab} \stackrel{\tilde g}{=}{\mathcal O}(\tilde s^k)\, ,
$$ 
for $k$ as high as possible. Since we are working with four-dimensional spacetimes, this problem is easily tractable.
In particular, since necessarily $\tilde n|_S=\hat{\tilde n}$ and $\mathring K$ extends $\IIo^{S\hookrightarrow \scri}$ to $M$, it follows that the left-hand side of the above display vanishes along $S$. 
Without loss of generality, we may assume that our initial extension obeys $\tilde n^a \mathring K_{ab}\stackrel{\tilde g}{=}0$ everywhere in $M$. Then,
in order that the above-displayed left-hand side is ${\mathcal O}(\tilde s^2)$, one needs only impose that
$$
%\tilde{n}^a \mathring{K}_{ab} =0, \quad  
\tilde {n}^a \mathring{K}^{(1)}_{ab} \stackrel S=  \tilde\nabla^a \mathring{K}_{ab}\, .
$$
Observe that this does not completely determine~$\mathring{K}^{(1)}_{ab}$ along $S$.
Again, without losing generality, we may assume that $\tilde n^a \mathring{K}^{(1)}_{ab}\stackrel{\tilde g}{=}  \tilde\nabla^a \mathring{K}_{ab}$ everywhere in $M$.
At this point we have that
\begin{equation} 
\tilde s\hh\tilde\nabla^a \IIo^{M, \textrm{as}}_{ab}-2\tilde n^a\IIo^{M, \textrm{as}}_{ab}\stackrel{\tilde g}{=}
\tilde s^2 L_b
+{\mathcal O}(\tilde s^3)\, ,
\end{equation}
where $L_a \stackrel S= \tilde \nabla^b \mathring{K}^{(1)}_{ab}$. 

The next order coefficient tensor~$\mathring{K}^{(2)}_{ab}$ in the expansion is critical for two reasons:
First, the tensor~$\mathring K^{(2)}|_S$ is completely undetermined and consists of both Neumann data for
the constraint system as well as initial data for the spacetime.
Second, thanks to our smoothness assumption (leading to the above  ansatz for the asymptotics of~$\IIo^M$), the 
 tensor $L_a$ must vanish  along~$S$. The case where $L_a|_S\neq 0$ and logarithms are needed is analyzed in  Section~\ref{notreg}.

We now seek a natural, conformally invariant tensor capturing Neumann data for the almost momentum constraint corresponding to  the tensor~$\mathring{K}^{(2)}$ in the expansion in Equation~\eqref{komasym}.
%Fortunately it is not difficult to construct such a tensor.
 We construct such a tensor and denote it by $p_2^{\rm ren}$, 
\begin{equation}\label{p2ren}
(p_{2}^{\rm ren})_a := \tilde{\bar{\top}} \big( \tfrac{1}{2} \hat{\tilde n}^b \hat{\tilde n}^c \hat{\tilde n} ^d \tilde \nabla_b \tilde \nabla_c \IIo_{da}^{M} + \tfrac{3}{2} H^{S \hookrightarrow M} \tilde{\bar{\nabla}}^b \IIo^{S \hookrightarrow \scri}_{ab} - \tfrac{1}{2} \IIo^{S \hookrightarrow \scri}_{ab} \tilde{\bar{\nabla}}^b H^{S \hookrightarrow M} \big)\big|_{S}.
\end{equation}
It probes the correct number of normal derivatives off of~$S$, is conformally invariant on a maximal Cauchy hypersurface $M$ (maximality ensures that
the  conformally compact metric $\tilde g^o$ is asymptotically hyperbolic and the embedding $S\hookrightarrow M$ is umbilic), and moreover
 \begin{equation} \label{ivp2ren}
 (p^{\textrm{ren}}_2)_a =\tilde{\bar{\top}} ( C_{\hat{\tilde n} \hat{\bar m} a})\, .
\end{equation}
This is proved in Appendix \ref{apenp2ren}.
 
 \section{Non-Willmore-flat Cauchy cuts}\label{notreg}

The Ashtekar--Das charge of Equation~\eqref{acharge} is well-defined for any Cauchy cut $S\hookrightarrow (\scri,\bar{\bm c})$, and in particular when this conformal embedding is not Willmore flat.
We now consider the case when the Willmore invariant ${\mathcal B}^{S\hookrightarrow \scri}$ is non-vanishing. This implies the presence of logarithms in the boundary asymptotics of the defining density for the maximal Cauchy hypersurface $M$~\cite{grahwit,gwa}, so we must reconsider the almost constraint Equations~\eqref{BP}.

The almost constraint equations extend the constraints on $\mathring M$ smoothly to the boundary so long as the conformal class induced on $M$ itself is smooth.
Our zeroth step is to examine the regularity of the conformal class of metrics $\tilde{\bm c}$ on $\mathring M$ obtained by pulling back the smooth conformal class $\bm c$ on~$X$. When the Willmore invariant ${\mathcal B}^{S\hookrightarrow \scri}$ is non-vanishing, smoothness of the defining density $\tau$ for $M\hookrightarrow X$ is obstructed at lowest order by terms of order $ s^4\log  s$; see \cite{gwa}. This would seem to imply the presence of terms of order $\tilde{s}^3\log \tilde s$ in the boundary asymptotics of the  pullback $\tilde g$ of the compactified metric $ g$  to $M=\tau^{-1}(0)$. However, using that $M$ intersects $\scri$ at right angles, it follows that the asymptotics of the pullback metric $\tilde g$ have lowest order logarithms  at order $\tilde{s}^4\log \tilde s$. This result can be found in~\cite{grar,chrgal}, but also follows from the analysis of~\cite{grahwit}.

We must now consider 
the constraint Equations~\eqref{BP} on
a Riemannian manifold $(M,\tilde g)$ where~$\tilde g$ 
has the non-smooth boundary asymptotics mentioned above.
Terms of order $\tilde s^4 \log \tilde s$ in $\tilde g$ 
contribute leading order terms  $\tilde s^4 \log \tilde s$ in the tensors 
$|\tilde n|^2_{\tilde g}+2 \tilde \rho \tilde{s} -1 +\frac16 \tilde s^2
|\IIo^{M}|^2_{\tilde g}$ and
 $\tilde s\tilde\nabla^a \IIo^{M}_{ab}-2\tilde n^a\IIo^{M}_{ab} $ appearing in the almost constraint Equations~\eqref{BP}. These   are of sufficiently high order that they do not impact the analysis that follows.

To analyze the almost Hamiltonian constraint, let us consider a smooth asymptotic solution 
$$\tilde{\sigma}_{\rm aHc}= \tilde \sigma_0(1+  A_1 \tilde \sigma_0+A_2 \tilde \sigma_0^2+\cdots)\, ,$$ given as an expansion in powers of some smooth defining density $\tilde \sigma_0$, where the $\cdots$'s denote  smooth terms involving higher powers of $\tilde\sigma_0$ that one attempts to determine to the highest order possible.  It follows from a simple generalization of the analysis of~\cite{ACF, will} that this asymptotic problem is already obstructed by a cubic order term $\tilde\sigma_{\rm aHc}^3 B
$, appearing in the Hamiltonian constraint equation evaluated on $\tilde \sigma_\mathrm{aHc}$, which cannot be removed without including logarithms in the asymptotics. The failure of~${ B}|_S$ to vanish is the obstruction to smoothness of the solution $\tilde{\sigma}$.
Recall that the obstruction to the  singular Yamabe problem for the conformal embedding $S\hookrightarrow (M,\tilde{\bm c})$ is the Willmore invariant~${\mathcal B}^{S\hookrightarrow M}$ given by Equation~\eqref{Willmore}; see~\cite{ACF,gwln,will}. 
We here treat the analogous prescribed scalar curvature problem
coming from the Hamiltonian constraint. 
 By applying the recursion given in~\cite[Section 3]{will} we find
$$
B|_S\propto {\mathcal B}^{S\hookrightarrow M}-\tfrac{1}{2} (\tilde \nabla_{\hat{\tilde n}} +2H^{S\hookrightarrow M})  |\IIo^{M}|^2_{\tilde g}\big|_S\, .
$$ 
Note that the  operator acting on the rigidity density~$|\IIo^{M}|^2_{\tilde g}$ is proportional to the conformally invariant Robin operator of~\cite{Cherrier}.

We can use the condition that $M\hookrightarrow(X,g^o)$ is a maximal embedding 
to further analyze the obstruction~$B|_S$. Recall from Equation~\eqref{iism} that umbilicity of the conformal embedding $\scri \hookrightarrow(X,\bm c)$ implies the same of $S\hookrightarrow(M,\tilde{\bm c})$, and thus from Equation~\eqref{Willmore} we see that $ {\mathcal B}^{S\hookrightarrow M}=0$. 
Note that, as implied by the discussion in Section~\ref{Constraints},
no logarithms can appear in the asymptotics of~$\IIo^M$ before order $\tilde s^2 \log \tilde s$ so the Robin operator acting on the rigidity density~~$|\IIo^{M}|^2_{\tilde g}$ of $M\hookrightarrow (X,\bm c)$  is well-defined.
In fact this vanishes 
as can be established by differentiating the almost maximal Equation~\eqref{am2as} twice and then applying the Codazzi--Mainardi identity in Equation~\eqref{CM}. Hence we have that~$B|_S=0$.

For $h_4^{\textrm{ren}}$ to be well-defined, we need to take four normal derivatives of $\tilde s$. 
Therefore we must check at which order logarithms in $|\IIo^M|^2_{\tilde g}$ impact the asymptotics of $\tilde s$. Since that term is
multiplied by~$\tilde s^2$ as it appears  in the almost Hamiltonian constraint of Equation~\eqref{BP}, its first non-smooth contribution there 
occurs only at order~$\tilde s^4\log\tilde s$. 
Following the recursive argument of~\cite[Section~4.3]{gwln}, these terms require logarithmic corrections to the defining density
only at order $\tilde{s}^5\log \tilde{s}$, which combined with the above-discussed regularity of $\tilde g$ implies that  $h_4^{\textrm{ren}}$ is  well-defined; see Appendix~\ref{apenmain} for details.

\medskip

We now consider the almost momentum constraint. The obstruction $\bar L_a:=L_a|_S$ to smooth solutions reads
$$
\bar L_a \stackrel{\tilde g}{=}\tfrac12 \hat{\tilde n}^c \hat{\tilde n}^d\tilde \nabla_c \tilde \nabla_d \big(\tilde s\hh\tilde\nabla^b \IIo^{M,\textrm{as}}_{ab}-2\tilde n^b\IIo^{M, \textrm{as}}_{ab}\big)\big|_S
= - 6 \hat{\tilde n}_a  b|_{S}\propto \hat{\tilde n}_a {{\mathcal B}}^{S\hookrightarrow \scri}\, .
$$
The detailed proof of this relation is given in Appendix~\ref{apenwillmore}. Observe that it  follows that the same ${\mathcal B}^{S\hookrightarrow\scri}=0$  condition, which would imply that the embedding $M \hookrightarrow (X, \bm c)$ is smooth, also ensures that the asymptotics of $\IIo^M$ (as analyzed in Equation~\eqref{komasym}) are  smooth. 

Since the obstruction $\bar{L}_a$ to the smooth asymptotics of the
trace-free extrinsic curvature $\IIo^{M}_{ab}$ is
proportional to the Willmore invariant
$\mathcal B^{S\hookrightarrow \scri}$, we now have to include terms
with logarithms, so
\[
\tilde{n}^b \IIo^{M}_{ab}
\stackrel{\tilde g}=
\tilde{s}\tilde{\nabla}^b\mathring{K}_{ab}
+\tilde{s}^2\left(
\tilde{n}^b\mathring{K}^{(2)}_{ab}
-\log\tilde s\,L_a
\right)
+\mathcal{O}_{\rm phg}(\tilde{s}^3 \log \tilde s)\, .
\]
Here $\mathcal{O}_{\rm phg}$ allows for a polyhomogeneous expansion in powers of the logarithm $\log \tilde s$; see~\cite{chrand} for a study of polyhomogeneous
asymptotic solutions of the almost momentum constraint. Importantly, even in the presence of logarithmic terms, the second normal-order operator acting on
$\IIo^M$, as in Equation~\eqref{p2ren}, is still  well defined. This is proved in Appendix~\ref{apenmain}.

\section{Maximal Cauchy hypersurface charge formul\ae}\label{proofth1}
We are now ready to reexpress the Ashtekar--Das charge $Q_{\bar \xi}$ of Equation~\eqref{acharge}  as an invariant of an initial maximal Cauchy hypersurface~$M$, thereby proving 
Theorem~\ref{mainth}.
 First we decompose 
the conformal Killing vector ${\bar \xi}$ of  $(\scri,\bar{\bm c})$
 into its normal and projected parts along the Cauchy cut $S\hookrightarrow (\scri,\bar {\bm c})$, 
$$
{\bar \xi}^a \stackrel{S}{=} \nu \hat{\bar m} ^a  + \tilde{{\bar \xi}}^a\, ,
$$
where $\nu := - {\bar \xi} . \hat{\bar m} $ and $\tilde{{\bar \xi}} :=\tilde{\bar{\top}}( {\bar \xi})$. Along~$S$ we then  have,
$$
\IVo^{\scri} (\hat{\bar m}, {\bar \xi})  \stackrel{S}{=} \nu \ \IVo^{\scri}_{\hat{\bar{m}}\hat{\bar{m}}}  +  \tilde{{\bar \xi}}^a 
%\tilde{\bar \top} 
%\big(
\IVo^{\scri}_{\hat{\bar m} a}
%\big)
  \, .
$$ 
It only remains to express the above  projections of $\IVo^{\scri}_{\hat{\bar m}a}$ in terms of the geometry of $M \hookrightarrow X$ along~$S$ using the almost maximal equation for $M$. This computation is performed in Appendix~\ref{apenmain} in the most general case where the Cauchy cut~$S$ is not Willmore, and gives
\begin{equation}\label{4toh4p2}
\IVo^{\scri}_{\hat{\bar{m}} \hat{\bar{m}}} =  h^{\textrm{ren}}_4
\:\mbox{ and }\:
\tilde{\bar{\top}} (\IVo^{\scri}_{\hat{\bar{m}}  a})  = (p^{\textrm{ren}}_2)_a\, .
\end{equation}
This establishes that the Ashtekar--Das charge can be expressed as a boundary integral of the Neumann data for the constraint equations on a  spacetime Cauchy hypersurface
$$
Q_{\bar \xi}
%=-\int_{(S,\, \,  \tilde \bar {\!\!\bm c}) } \IVo^{\scri}(  \hat {\bar m},{\bar \xi})
 = -
\int_{(S,\, \,  \tilde \bar {\!\!\bm c}) } 
\big(
 \nu h^{\rm ren}_4 + p_2^{\rm ren}(\tilde{{\bar \xi}})
\big)\,  .
$$
Having proved Theorem~\ref{mainth}, we now turn to examples.

\subsection{Hamiltonian charges}
We  now compare  our  maximal Cauchy hypersurface  expression for the Ashtekar--Das charge 
with the 
Hamiltonian charges studied in~\cite{chrnag}. In that approach, one models spacetime initial data by a three-manifold~$M$ equipped with an extrinsic curvature field
and an asymptotically hyperbolic metric that approaches a
reference hyperbolic metric~$\tilde g_{\rm hyp}$ near the
conformal boundary $S=S^2$. We thus consider ans\"atze for the Cauchy hypersurface singular metric and extrinsic curvature given by
\begin{equation} \label{ahypid}
\tilde{g}^o =\tilde{s}^{-2} \left(\ext \tilde{s}^2 +\left( 1-\tfrac{\tilde{s}^2}{4} \right)^2 g_{S^2} + \tilde{s}^3 \tilde{g}^{(3)} + \mathcal{O}(\tilde{s}^4) \right) =: \tilde{g}_{\rm hyp} + \tilde{s} \tilde{g}^{(3)}  + \mathcal{O}(\tilde{s}^2), \qquad \IIo^{M}_{\tilde g^o} 
%\stackrel{\ \tilde{g}^o}{=} 
=
\tilde{s} \mathring{\kappa}\, .
\end{equation}
Note that we have written the metric in Graham--Lee normal form, where $\tilde s$ is the  geodesic distance  
to the boundary of the compactified metric $\tilde s^2 \tilde g^o$. Also, $g_{S^2}$ is the standard round sphere metric and~$\tilde g^{(3)}$ is a tangential symmetric two-tensor on the level sets of $\tilde s$. For simplicity we have taken the special metric asymptotics considered by Wang~\cite{wang}. We also assume that $\mathring{\kappa}$ extends smoothly to~$S$ so the above fall-off of the extrinsic curvature is two orders faster than the case handled by our general analysis. 
 This corresponds to setting
$\mathring K$ and $\mathring K^{(1)}$ to zero in Equation~\eqref{komasym}.

Hamiltonian charges correspond to reference symmetries. Therefore one considers the three-dimensional hyperbolic manifold defined by the $t=0$ slice of  anti-de Sitter space, in coordinates where the  global timelike Killing vector equals $\frac{\partial}{\partial t}$. The
 \emph{Killing initial data} 
corresponding to any anti-de Sitter Killing vector  ${ \xi}^{\textrm{AdS}}$ is then defined by the  pair~$(\alpha,\beta)$
consisting of the  
 normal and projected parts of ${\xi}^{\textrm{AdS}}$ with respect to the future-pointing timelike unit normal vector to the $t=0$ hypersurface. Global charges associated to the Killing initial data~$(\alpha,\beta)$ can now be defined on
  an asymptotically hyperbolic Cauchy hypersurface as follows~\cite{chrnag} (see also~\cite{michel,Cederbaum}),
\begin{equation} \label{chr}
Q_{(\alpha, \beta)} := 
- \lim_{\tilde{s} \rightarrow  0} 
\int \displaylimits_{S^2} 
\tilde{s}^{-1}\left(1-\tfrac{\tilde{s}^2}{4} \right)^2\hat{n} . \mathfrak{U}_{(\alpha, \beta)} (\gamma, \II^M )  \  \mathrm{d}\mathrm{Vol}_{g_{S^2}}\, ,
\end{equation}
where  $\hat{n}=\partial_{\tilde{s}}$, $\gamma:=\tilde{g}^o-\tilde{g}_{\rm hyp}$, $\II^M$ is the extrinsic curvature of  $\mathring M$ with respect to $g^o$, and
\begin{align*}
\mathfrak{U}_{(\alpha, \beta)} (\gamma, \II^M )&:=  \alpha \left( \rm{div}_{\tilde{g}_{\rm hyp}} \gamma- \ext \left( \rm{tr}_{\tilde{g}_{\rm hyp}} \gamma \right) \right) -  \gamma(\nabla^{\sharp}_{\tilde{g}_{\rm hyp}} \alpha, \pdot) + \rm{tr}_{\tilde{g}_{\rm hyp}} \gamma \ \ext \alpha \\
& \  +2 \left(\II^M (\beta, \pdot) - \rm{tr}_{\tilde{g}_{\rm hyp}} \II^M  \  \tilde{g}_{\rm hyp} (\beta, \pdot) \right).
\end{align*}
We use the Killing initial data basis of \cite[Section 4]{Cederbaum}, expressed in the geodesic coordinate $\tilde s$, and adopt the terminology of \cite{chrmt} for the corresponding charges:
\begin{itemize}
\item  {\it Energy momentum}: $Q_{(\alpha_{(\mu)},0)}$ 
with
$
\alpha_{(0)} := \tfrac{1}{\tilde{s}} + \tfrac{\tilde{s}}{4}$,  $\alpha_{(i)} :=\left( \tfrac{1}{\tilde{s}} - \tfrac{\tilde{s}}{4} \right) \omega^i\, ,
$\\[.5mm]
\item
 {\it Center-of-mass}:  
$Q_{(0, \beta_{(i)})}$ with
$
\beta_{(i)} := \tilde{s} \omega^i \partial_{\tilde{s}} - \tfrac{1+\tilde{s}^2/4}{1-\tilde{s}^2/4} \operatorname{grad}_{g_{S^2}} \omega^i\, ,
$\\[.5mm]
\item 
{\it Angular momentum}: 
$Q_{(0, J_{(i)(j)})}$ 
with
$
J_{(i) (j) } :=2  \omega^{[i}
\operatorname{grad}_{g_{S^2}}
\omega^{j]}\,.
$
\end{itemize}
The three functions $\omega^i\in C^\infty(S^2)$ in the above are the pullbacks of the standard Cartesian coordinates in~${\mathbb R}^3$ to a unit sphere, so obey  $\sum_{i=1}^3 (\omega^i)^2 =1$ as well as the Hessian condition
 $\nabla_{g_{S^2}} \nabla_{g_{S^2}} \omega^i  = - \omega^i g_{S^2} $.
Computing these charges for the asymptotically hyperbolic initial data in Equation~\eqref{ahypid} gives 
$$
Q_{(\alpha_{(\mu)},0)} = 3\big \langle \omega^{\mu}  \rm{tr}_{g_{S^2}} (\tilde{g}^{(3)}|_{S})\big \rangle , \quad Q_{(0, \beta_{(i)})} = 2 \big\langle \mathring{\kappa} (\partial_{\tilde{s} },\operatorname{grad}_{g_{S^2}}\omega^i )  \big\rangle, \quad Q_{(0, J_{(i)(j)})} = -2 \big\langle  \mathring{\kappa} (\partial_{\tilde{s}}, J_{(i)(j)})   \big\rangle,
$$
where $\omega^0 = 1$ and $\langle f \rangle := \int_{S^2} f  \ \rm{dVol}_{g_{S^2}}$.  

\medskip

We now compute  Ashtekar--Das charges using the  Neumann data for the almost constraint equations. Note that   the conformal boundary $S$ for the metric representative given in Equation~\eqref{ahypid} is {totally} geodesic.
Because $\tilde s$ is a Gaussian normal coordinate, 
$\hat{n}^a \hat{n}^b \hat{n}^c \hat{n}^d \tilde{\nabla}_a  \tilde{\nabla}_b  \tilde{\nabla}_c \tilde{\nabla}_d \tilde{s}$,
  $\hat{n}^a \hat{n}^b \hat{n}^c \tilde{\nabla}_a   \tilde{\nabla}_b \tilde{\nabla}_c \tilde{s}$ and $\hat{n}^a \hat{n}^b  \tilde{\nabla}_a  \tilde{\nabla}_b  \tilde{s}  $ all vanish along $S$. Thus, using Equations~\eqref{h4ren} and~\eqref{p2ren}, 
$$
h_4^{\rm ren} =  \hat{\tilde n}^a \hat{\tilde n}^b \hat{\tilde n}^c \tilde{\nabla}_a \tilde{P}_{bc} \big|_S\:
\mbox{ and }\: p_2^{\rm ren}  = \tilde{\bar{\top}} \mathring{\kappa} (\hat{\tilde n}, \cdot ) \big|_S\, ,
$$
where $\IIo^M \stackrel{\tilde{g}} = \tilde{s}^2 \mathring{\kappa}$ has been used in the second equality. Using the form of the metric from Equation~\eqref{ahypid}, we can show that
$ \hat{\tilde n}^a \hat{\tilde n}^b \hat{\tilde n}^c \tilde{\nabla}_a \tilde{P}_{bc} \big|_S=- \tfrac{3}{2}\rm{tr}_{\tilde{\bar{g}}} (\tilde{g}^{(3)}|_{S})$, so 
$$
Q_{\bar \xi} = 
\int_{(S,\, \,  \tilde \bar {\!\!\bm c}) } 
\big(
 \tfrac{3}{2} \nu\hh  \rm{tr}_{\tilde{\bar{g}}}( \tilde{g}^{(3)})-  \mathring{\kappa} (\hat{\tilde n}, \tilde{{\bar \xi}} )
\big)\,  .
$$
Now suppose that  the spacetime conformal boundary is the Einstein cylinder $(\mathbb{R}\times S^2, [-\mathrm{d}t^2+g_{S^2}] )$, with $S\cong S^2$ corresponding to the cut $t=0$. The boundary conformal Killing vectors induced by the anti-de Sitter Killing vectors restrict, in the notation used above, to $S$ as follows
$$
{\bar \xi}_{(0)} = \partial_t, \quad {\bar \xi}^{+}_{(i)} = \omega^i \partial_t,\quad  {\bar \xi}^{-}_{(i)} = - \operatorname{grad}_{g_{S^2}} \omega^i, \quad {\bar \xi}_{(i)(j)} = 2 \omega^{[i}\operatorname{grad}_{g_{S^2}} \omega^{j]}\, .
$$
The energy-momentum vector corresponds to evaluating $Q_{{\bar \xi}}$ for $({\bar \xi}_{(0)}, {\bar \xi}^{+}_{(i)})$, the center-of-mass components to~${\bar \xi}^{-}_{(i)}$ and the angular momentum to~${\bar \xi}_{(i)(j)}$. For each corresponding pair of reference Killing initial data~$(\alpha,\beta)$ and a boundary conformal Killing field ${\bar \xi}$, the
preceding expressions give
$
Q_{(\alpha,\beta)}=2Q_{\bar \xi}
$.
Restoring the respective conventional four-dimensional prefactors
$1/(16\pi G)$ and~$1/(8\pi G)$,  the Hamiltonian and
Ashtekar--Das charges agree precisely.

\section*{Acknowledgements}
A.R.G. and A.W. acknowledge support from the Royal Society of New Zealand via Marsden Grant 24-UOA-005. A.W. was also supported by Simons Foundation Collaboration Grant for Mathematicians ID 686131. J.K. acknowledges support from the National Science Centre, Poland, under the SONATA project No. 2025/59/D/ST2/02050. This work was partially supported by the Simons Foundation grant SFI-MPS-T-Institutes-00010825, and by the State Treasury funds as part of a task commissioned by the Minister of Science and Higher Education under the project Organization of the Simons Semesters at the Banach Center - New Energies in 2026--2028 (MNiSW/2025/DAP/491). 

\appendix

\section{Computation of the Hamiltonian constraint Neumann-type data} \label{apenh4}
Here we derive the expression given in
Equation~\eqref{pi14} for the Hamiltonian constraint Neumann-type data $h_4^{\rm ren}$.
We work in the setting of Section~\ref{Constraints}, with
$\tilde s=s|_M$. Throughout we choose some $g\in \bm c$ and  use the gradient
extensions $n=\ext s$ and $m=\ext t$ to the ambient space $X$ of the unit conormals of $\scri$ and $M$; see Equations~\eqref{aeeq} and~\eqref{asymyt}. Their restrictions to $S$ are $\hat{\tilde n}$ and
$\hat{\bar m}$, respectively. The tangential projector from $TX$ to~$TM$ is
$\tilde{\top}_a{}^b=\delta_a{}^b+\hat m_a\hat m^b$, so that
\[
\tilde n_a:=\tilde\nabla_a\tilde s
=\tilde{\top}_a{}^b n_b\big|_M
=\left(\delta_a{}^b-\tfrac{m_a m^b}{g(m,m)}\right)n_b\big|_M.
\]
We need to compute  higher derivatives $\tilde \nabla^k\tilde s$. For that we express these in terms of $\nabla$, $n$ and $m$ in order to use the almost Einstein, maximal, and asymptotic singular Yamabe Equations~\eqref{aeeq}, \eqref{AMC}, and~\eqref{asymyt}.

\smallskip

The almost Einstein equation and the singular Yamabe normalization give (see for example~\cite{will})
\begin{equation}\label{apengradientjets}
\nabla_a n_b\big|_{\scri}
\stackrel{g}{=}H^{\scri}g_{ab}\big|_{\scri}\, ,\qquad
\nabla_a m_b\big|_M
\stackrel{g}{=}\big(\IIo^M_{ab}+H^M g_{ab}\big)\big|_M\, .
\end{equation}
Projecting the first relation to directions tangent to $M$ and using that $H^{S \hookrightarrow M}  = H^{\scri} \big|_{S}$, shows that  when the Hessian of $\tilde s$ along $M$  is
 restricted to  $S$, one has 
\begin{equation}\label{apengradientjets1}
\tilde\nabla_a\tilde n_b\big|_S
\stackrel{\tilde g}{=}H^{S\hookrightarrow M}\tilde g_{ab}\big|_S\, .
\end{equation}
Also, Equation~\eqref{mn0} implies that the extensions $m$ and $n$ obey
$
n.m\big|_S=0\ .
$

To compute $h_4^{\rm ren}$, we first need to compute
$\tilde \pi_1$, $\tilde \pi_2$ and $\tilde \pi_3$---as given in Equations~\eqref{pi1}, \eqref{pi2} and~\eqref{pi3}---when  evaluated on the defining density 
$\tilde \sigma=[\tilde g,\tilde s]$.
The first two boundary curvature scalars are therefore
\begin{align}
\tilde\pi_1
&\stackrel{\tilde g}{=}\hat{\tilde n}^a\tilde{\top}_a{}^b n_b\big|_S=1\, ,\label{Pi1}\\
\tilde\pi_2
&\stackrel{\tilde g}{=}\big(\hat{\tilde n}^a\hat{\tilde n}^b\tilde\nabla_a\tilde\nabla_b\tilde s
-H^{S\hookrightarrow M}\big)\big|_S
=H^{\scri}\big|_S-H^{S\hookrightarrow M}=0\, .\label{Pi2}
\end{align}
The second relation above used Equation~\eqref{apengradientjets1}.
For $\tilde \pi_3$, differentiating the Hessian of $\tilde s$ once more gives
\begin{equation}\label{Pi3}
\tilde\pi_3
=\big((H^M)^2-2P_{\hat{\tilde n}\hat{\tilde n}}
+2\tilde P_{\hat{\tilde n}\hat{\tilde n}}\big)\big|_S.
\end{equation}
Here we used the second consequence of the almost Einstein equation listed in Equation~\eqref{almE} below.
Since the ambient Weyl tensor vanishes on $\scri$, the normal-normal component of the Gauss Equation~\eqref{Fialkow} applied to
$M\hookrightarrow X$, together with $\hat{\tilde n}^b\IIo^M_{ab}|_S=0$, yields
\begin{equation}\label{pnnw}
P_{\hat{\tilde n}\hat{\tilde n}}\big|_S
=\big(\tilde P_{\hat{\tilde n}\hat{\tilde n}}
+\tfrac12(H^M)^2+\tfrac14|\IIo^M|_{\tilde g}^2\big)\big|_S\, .
\end{equation}
Consequently,
\[
\tilde\pi_3=-\tfrac12|\IIo^M|_{\tilde g}^2\big|_S\, .
\]

We are now ready to compute $h_4^{\rm ren}$. 
Since $S\hookrightarrow M$ is umbilic
and $\tilde\pi_1=1$, $\tilde\pi_2=0$, Equation~\eqref{h4ren} becomes
\begin{equation}\label{h4renappen}
h_4^{\rm ren}\stackrel{\tilde g}{=}\tfrac13\big[
\hat{\tilde n}^a\hat{\tilde n}^b\hat{\tilde n}^c\hat{\tilde n}^d
\tilde\nabla_a\tilde\nabla_b\tilde\nabla_c\tilde\nabla_d\tilde s
+H^{\scri}\big(2\tilde P_{\hat{\tilde n}\hat{\tilde n}}
-|\IIo^M|_{\tilde g}^2\big)+3\hat{\tilde n}^a\hat{\tilde n}^b\hat{\tilde n}^c\tilde\nabla_a\tilde P_{bc}
\big]\big|_S\, .
\end{equation}
Now we list some main identities needed to further develop the above expression. 
First 
we note that the fully normal contraction of the first covariant derivative of $\IIo^M$ vanishes along $S$:
\begin{equation}\label{apennormalKjet}
\hat{\tilde n}^a\hat{\tilde n}^b\hat{\tilde n}^c
\tilde\nabla_a\IIo^M_{bc}\big|_S=0\, .
\end{equation}
The above identity can be derived in the following way. Differentiate the almost momentum constraint in
Equation~\eqref{almcon} once in the normal direction and then use
Equation~\eqref{apengradientjets}. Then consider the  normal component and further
split the divergence into normal and surface contributions.
Tangential differentiation of the vanishing normal contraction
of $\IIo^M$ shows that the surface contribution is proportional
to the trace of its pullback to $S$, and therefore vanishes. The result then follows.

Next, restricting the almost Einstein equation to \(\scri\) and comparing its tangential part with the second fundamental form gives the boundary value of \(\rho\). Identities for its derivatives  follow by taking a divergence of the almost Einstein Equation, commuting covariant derivatives and using the contracted Bianchi identity. Differentiating the resulting bulk relation once more gives the  normal components of the Hessian of $\rho$ (see~\cite{gwln} and~\cite{will} for details)
\begin{align}\label{almE}
\rho\big|_{\scri}=-H^{\scri},\qquad
\nabla_{\hat n}\rho\big|_{\scri}=P_{\hat n\hat n}\big|_{\scri}, \quad \hat n^a\hat n^b\nabla_a\nabla_b\rho\big|_{\scri}
=\big(\hat n^a\hat n^b\hat n^c\nabla_aP_{bc}
+H^{\scri}P_{\hat n\hat n}\big)\big|_{\scri}.
\end{align}
Computing two normal derivatives of the almost Einstein equation and contracting with two further normal vectors gives
\begin{equation}\label{amb4}
\big(\hat n^a\hat n^b\hat n^c\hat n^d
\nabla_a\nabla_b\nabla_c n_d
+2H^{\scri}P_{\hat n\hat n}
+3\hat n^a\hat n^b\hat n^c\nabla_aP_{bc}\big)\big|_{\scri}=0\, .
\end{equation}
The above can also be derived using that the boundary curvature scalar $\pi_4$ of the spacetime defining density $\sigma$ must vanish; see~\cite{bcs}.
To use the above display to compute the first term on the right-hand side of Equation~\eqref{h4renappen} requires handling  
derivatives of an extension of the projector $\tilde \top$. In particular we need the identity 
\begin{equation}\label{jet2}
\hat n^a\hat n^b\nabla_a\nabla_b(n_c m^c)\big|_S
\stackrel{g}{=}2\Big(\big(\bar\nabla_{\hat{\bar m}}+H^M 
\big)H^{\scri}\Big)\Big|_S.
\end{equation}
This can be obtained by
differentiating Equation~\eqref{am2} twice and then using the extrinsic curvature identities given at the end of Section~\ref{MCH}.

We still need to handle one normal derivative of both Schouten tensors as well as the Laplacian of mean curvature:
\begin{align*}
\hat n^a\hat n^b\hat n^c\nabla_aP_{bc}\big|_{\scri}
&=\big(\nabla_{\hat n}J-\tfrac32(H^{\scri})^3
-3H^{\scri}P_{\hat n\hat n}
+H^{\scri}\bar J
+\bar\Delta H^{\scri}\big)\big|_{\scri}\, ,\\
\hat{\tilde n}^a\hat{\tilde n}^b\hat{\tilde n}^c\tilde\nabla_a\tilde P_{bc}\big|_S
&=\big(\tilde\nabla_{\hat{\tilde n}}\tilde J
+H^{\scri}\tilde J
-3H^{\scri}\tilde P_{\hat{\tilde n}\hat{\tilde n}}
+\tilde{\bar\Delta}H^{\scri}\big)\big|_S\, ,\\
\bar\Delta H^{\scri}\big|_S
&=\big(\tilde{\bar\Delta}H^{\scri}
-\hat{\bar m}^a\hat{\bar m}^b\bar\nabla_a\bar\nabla_bH^{\scri}
-2H^M\bar\nabla_{\hat{\bar m}}H^{\scri}\big)\big|_S\, .
\end{align*}
The above used the Gauss and
Codazzi--Mainardi equations and the Bianchi identity $\nabla_a J = \nabla^b P_{ab}$.
We have also used  
$\hat n^a\hat n_b=\delta^a{}_b-\bar\top^a{}_b$, as well as that
$\IIo^{\scri}$ and $\IIIo^{\scri}$ both vanish.

We now assemble the fourth boundary curvature scalar in three steps. 
We shall use Equation~\eqref{amb4} 
to compute
$
\hat n^a 
\hat n^b 
\hat n^c 
\hat n^d
\tilde \nabla_a
\tilde \nabla_b
 \tilde\nabla_c 
\tilde n_d
$
by expressing each $\tilde \nabla$ in terms of $\nabla+m\nabla_m$ working from the left.
This requires using
Equations~\eqref{apengradientjets} and ~\eqref{apennormalKjet} as well as the second normal jet of $n.m$ given in Equation~\eqref{jet2}. In doing so one also encounters the normal jets of $\rho$ given in Equation~\eqref{almE}.
 The second step is to handle the lower transverse order terms in
Equation~\eqref{h4renappen} by using Equation~\eqref{pnnw} for   the normal-normal component of the
ambient Schouten tensor. The final step is to substitute the results of the above display. 
Happily terms involving $\tilde{\bar\Delta}H^{\scri}$ then cancel.
Collecting the remaining terms and dividing
by three, we obtain
\begin{equation}\label{h4expanded}
\begin{split}
h_4^{\rm ren}=\Big[&
\tilde\nabla_{\hat{\tilde n}}\tilde J-\nabla_{\hat n}J
+5H^M\bar\nabla_{\hat{\bar m}}H^{\scri}
+\hat{\bar m}^a\hat{\bar m}^b\bar\nabla_a\bar\nabla_bH^{\scri}\\
&+H^{\scri}\big(\tfrac32(H^M)^2
+\tfrac32(H^{\scri})^2
+\tfrac14|\IIo^M|_{\tilde g}^2+\tilde J-\bar J\big)
\Big]\Big|_S\, .
\end{split}
\end{equation}

\smallskip
We still need to relate this expression to the normal component of the
fourth fundamental form. For that we first collect some necessary identities: The scalar Gauss Equation~\eqref{Ricci} applied to  $M$ and~$\scri$,
as well as the scalar Codazzi--Mainardi Equation~\eqref{SCM} for $\scri$, give
\begin{equation}  \label{pident}
\begin{aligned}
P_{ mm}\big|_M
&=\big(\tfrac32(H^M)^2-\tfrac14|\IIo^M|_{\tilde g}^2
-J+\tilde J\big)\big|_M,\\
P_{na}\big|_{\scri}
&=\big(\hat n_a P_{\hat n\hat n}
-\bar\nabla_aH^{\scri}\big)\big|_{\scri},\\
P_{n n}\big|_{\scri}
&=\big(\tfrac32(H^{\scri})^2+J-\bar J\big)\big|_{\scri}.
\end{aligned}
\end{equation}
We will also need the normal derivative of the rigidity density,
\[
\big(\tilde\nabla_{\hat{\tilde n}}
+2H^{\scri}\big)
|\IIo^M|_{\tilde g}^2\:\stackrel S=0\, .
\]
This follows from the Codazzi--Mainardi equation, vanishing of the ambient Weyl
tensor on $\scri$, and the almost momentum constraint.

Now 
we employ Equation~\eqref{ivo} to write
\begin{equation*}
\IVo^{\scri}_{\hat{\bar m}\hat{\bar  m}}\stackrel S=
C_{\hat{\tilde n}\hat{\bar m}\hat{\bar m}}\stackrel S
=n^a\nabla_a(P_{bc}m^b m^c)
-m^b m^c\nabla_b(P_{ac}n^a)-n^aP_{bc}\nabla_a(m^b m^c)
+m^b m^cP_{ac}\nabla_b n^a\, .
\end{equation*}
We can handle
derivatives of the extensions of $m$ and $n$ using Equation~\eqref{apengradientjets}. 
Then employing
Equation~\eqref{pident}
gives exactly the right-hand side of Equation~\eqref{h4expanded}.
Thus, 
\begin{equation}\label{h4renA}
h_4^{\rm ren}
=C_{\hat{\tilde n}\hat{\bar m}\hat{\bar m}}\big|_S
=\IVo^{\scri}_{\hat{\bar m}\hat{\bar m}}\big|_S\, ,
\end{equation}
which completes the proof of the identities in~\eqref{pi14}.

\section{ Proof of Equation~\eqref{ivp2ren} and conformal invariance of
$p_2^{\NoCaseChange{\mathrm{ren}}}$} \label{apenp2ren}

We will heavily rely on identities given in 
Appendix~\ref{apenh4}. In particular recall that
$H^{S\hookrightarrow M}=H^{\scri}|_S$, the pullback of $\IIo^M$
to $S$ equals $\IIo^{S\hookrightarrow\scri}$, and 
the almost momentum constraint of Equation~\eqref{almcon} implies  $\hat{\tilde n}^b\IIo^M_{ba}|_S=0$. 

We first record the normal  jets  of $\IIo^M$ needed for the calculation. 
One  normal derivative of the second identity of
Equation~\eqref{almcon}, along with
$\tilde\nabla_a\tilde n_b|_S=H^{S\hookrightarrow M}\tilde g_{ab}|_S$, implies that
\begin{equation}\label{p2firstjets}
\tilde{\bar{\top}}\big(
\hat{\tilde n}^b\hat{\tilde n}^c\tilde\nabla_b\IIo^M_{ca}\big)\big|_S
=\tfrac12\tilde{\bar{\top}}\big(
\tilde\nabla^b\IIo^M_{ba}\big)\big|_S=\tilde{\bar\nabla}^b\IIo^{S\hookrightarrow\scri}_{ab}\: \mbox{ and }\:\:
\hat{\tilde n}^a\hat{\tilde n}^b\hat{\tilde n}^c\tilde\nabla_a\IIo^M_{bc}\big|_S=0\, .
\end{equation}
To obtain the second equality, we split the divergence into its normal and
surface parts and reapplied the first equality. The last identity is the normal first-jet relation already recorded in Equation~\eqref{apennormalKjet}.

We are now ready to express  tangential Cotton components, and in turn those of the fourth fundamental form, in  terms of the given data along~$M$.
Equation~\eqref{ivo}  gives
\begin{align*}
\tilde{\bar{\top}}\big(\IVo^{\scri}_{\hat{\bar m}a}\big)\stackrel S
=\tilde{\bar{\top}}\big(&
\nabla_n P_{ma}-\nabla_m P_{na}
-P_{ca}\nabla_n m^c+P_{ba}\nabla_m n^b
\big)\, .
\end{align*}
The Codazzi--Mainardi relation for 
$
M\hookrightarrow X$ yields
\begin{equation}\label{CMMX}
\tilde{\top}(P_{ma})\big|_M
=\big(\tfrac12\tilde\nabla^b\IIo^M_{ab}
-\tilde\nabla_aH^M\big)\big|_M,
\end{equation}
while  the corresponding formula for $P_{na}|_\scri$  is given in
Equation~\eqref{pident}. In light of the above two displays, we will also need to relate  Hessians of the two mean curvatures, for which we differentiate the relation
$\tilde\nabla_{\hat{\tilde n}}H^M|_S=
\bar\nabla_{\hat{\bar m}}H^{\scri}|_S$ in directions tangential to $S$, giving
\begin{equation*}
\tilde{\bar{\top}}\big(
\hat{\tilde n}^b\tilde\nabla_b\tilde\nabla_aH^M\big)\big|_S
={}\tilde{\bar{\top}}\big(
\hat{\bar m}^b\bar\nabla_b\bar\nabla_aH^{\scri}\big)\big|_S+\IIo^{S\hookrightarrow\scri}_{ab}
\tilde{\bar\nabla}^bH^{\scri}
+H^M\tilde{\bar\nabla}_aH^{\scri}
-H^{\scri}\tilde{\bar\nabla}_aH^M.
\end{equation*}
Using this identity, the normal derivative formulas in
Equation~\eqref{pident} as well as Equation~\eqref{p2firstjets}, the above Cotton
components become
\begin{equation}\label{p2cotton}
\tilde{\bar{\top}}\big(\IVo^{\scri}_{\hat{\bar m}a}\big)\big|_S
={}\tfrac12\tilde{\bar{\top}}\big(
\hat{\tilde n}^b\tilde\nabla_b\tilde\nabla^c\IIo^M_{ca}\big)\big|_S+H^{S\hookrightarrow M}
\tilde{\bar\nabla}^b\IIo^{S\hookrightarrow\scri}_{ab}
-\IIo^{S\hookrightarrow\scri}_{ab}
\tilde{\bar\nabla}^bH^{S\hookrightarrow M}.
\end{equation}
The remaining step is to split the first term into its normal and
surface parts. This requires a second normal derivative of the
almost momentum constraint. 
Two normal derivatives of
$\tilde n^d\IIo^M_{da}$ 
can be computed in two distinct ways:
\begin{align*}
\tilde{\bar{\top}}\big(
\hat{\tilde n}^b\hat{\tilde n}^c\tilde\nabla_b\tilde\nabla_c
(\tilde n^d\IIo^M_{da})\big)\big|_S
={}&\tilde{\bar{\top}}\big(
\hat{\tilde n}^b\hat{\tilde n}^c\hat{\tilde n}^d
\tilde\nabla_b\tilde\nabla_c\IIo^M_{da}\big)\big|_S+H^{S\hookrightarrow M}
\tilde{\bar{\top}}\big(\tilde\nabla^b\IIo^M_{ba}\big)\big|_S
+\IIo^{S\hookrightarrow\scri}_{ab}
\tilde{\bar\nabla}^bH^{S\hookrightarrow M}\\
={}&\tilde{\bar{\top}}\big(
\hat{\tilde n}^b\hat{\tilde n}^c\hat{\tilde n}^d
\tilde\nabla_b\tilde\nabla_c\IIo^M_{da}\big)\big|_S+\tilde{\bar{\top}}\big(
\hat{\tilde n}^b\tilde{\bar g}^{cd}
\tilde\nabla_b\tilde\nabla_d\IIo^M_{ca}\big)\big|_S
+\tfrac12H^{S\hookrightarrow M}
\tilde{\bar{\top}}\big(\tilde\nabla^b\IIo^M_{ba}\big)\big|_S.
\end{align*}
The first line above was obtained by computing successive derivatives of $\tilde n^d$ using Equation~\eqref{apengradientjets1} followed by Equation~\eqref{p2firstjets}
to handle one derivative of $\tilde n^d$ as well as the almost Hamiltonian equation to handle two upon noting that $\tilde\nabla_{\tilde n} \tilde n^d = \frac 12 \tilde \nabla^d |\tilde n|^2_{\tilde g}$,
 while the second line employed the almost momentum constraint to replace $\tilde n^d\IIo^M_{da}$ by $\frac12\tilde s \tilde \nabla^b \IIo^M_{ba}$.
Using Equation~\eqref{p2firstjets} then gives
\begin{equation}\label{p2mixedderiv}
\tilde{\bar{\top}}\big(
\hat{\tilde n}^b\tilde{\bar g}^{cd}
\tilde\nabla_b\tilde\nabla_d\IIo^M_{ca}\big)\big|_S
=H^{S\hookrightarrow M}
\tilde{\bar\nabla}^b\IIo^{S\hookrightarrow\scri}_{ab}
+\IIo^{S\hookrightarrow\scri}_{ab}
\tilde{\bar\nabla}^bH^{S\hookrightarrow M}\, .
%=\tilde{\bar\nabla}^b\big(
%H^{S\hookrightarrow M}\IIo^{S\hookrightarrow\scri}_{ab}\big).
\end{equation}
Finally, writing 
$\big(\hat{\tilde n}^b\tilde\nabla_b\tilde\nabla^c\IIo^M_{ca}\big)\big|_S=\big((\tilde{\bar g}^{cd}+\hat{\tilde n}^c\hat{\tilde n}^d) 
\hat{\tilde n}^b\tilde\nabla_b\tilde\nabla_c\IIo^M_{da}\big)\big|_S
$
 in the first term of
Equation~\eqref{p2cotton} and then using Equations~\eqref{p2mixedderiv} and~\eqref{p2ren}
gives
\begin{align*}
\tilde{\bar{\top}}\big(\IVo^{\scri}_{\hat{\bar m}a}\big)\big|_S
={}&\tfrac12\tilde{\bar{\top}}\big(
\hat{\tilde n}^b\hat{\tilde n}^c\hat{\tilde n}^d
\tilde\nabla_b\tilde\nabla_c\IIo^M_{da}\big)\big|_S+\tfrac32H^{S\hookrightarrow M}
\tilde{\bar\nabla}^b\IIo^{S\hookrightarrow\scri}_{ab}
-\tfrac12\IIo^{S\hookrightarrow\scri}_{ab}
\tilde{\bar\nabla}^bH^{S\hookrightarrow M},
\end{align*}
which is $(p_2^{\rm ren})_a$, as required by Equation~\eqref{ivp2ren}. 
The left-hand side of the above display is
a conformally invariant one-form of weight $-2$, so too is the right-hand side. Finally, note when comparing the above to the formula appearing in Theorem~\ref{mainth}, one needs the first identity in Equation~\eqref{iism2}.

\section{The Willmore obstruction to the almost momentum constraint}
\label{apenwillmore}
We now work on the maximal hypersurface $M$ in the case where smoothness is obstructed.  Quantities along $M$ then have the 
polyhomogeneous boundary expansions discussed in Section \ref{notreg}: In any smooth ambient scale the first possible logarithms in the asymptotics of the defining function $t$ for~$M$, induced  metric~$\tilde g$, and
$\IIo^M$ have respective orders $ s^4\log s$,
$\tilde s^4\log\tilde s$, and $\tilde s^2\log\tilde s$.

\smallskip
As explained in Sections~\ref{Constraints} and~\ref{notreg} the smooth asymptotics 
$\IIo^{M,\mathrm{as}}$ of  the  solution $\IIo^M$ to the almost momentum constraint  obey
\begin{equation}\label{aresidual}
 \tilde s\,\tilde\nabla^b\IIo^{M,\mathrm{as}}_{ab}
 -2\tilde n^b\IIo^{M,\mathrm{as}}_{ab}
 \stackrel{\tilde{g}}{=}\tilde s^2L_a+\mathcal O_{\rm phg}(\tilde s^3),
\end{equation}
where  $ \bar L_a=L_a|_S$ is the obstruction to smoothness of $\IIo^M$. 
Recall also that $\IIo^{M,\mathrm{as}}+\tilde s^2 A$ also solves the above equation for any choice of smooth tensor $A$ because $\tilde s\,\tilde\nabla^b(\tilde s^2 A_{ab})
 -2\tilde n^b\tilde s^2 A_{ab}
\stackrel{\tilde{g}}{=}\tilde s^3\tilde\nabla^bA_{ab}$. Because $M$ is not smooth to the boundary, the remainder $\mathcal O_{\rm phg}(\tilde s^3)$ on the right-hand side is a polyhomogeneous expansion.

Now consider an asymptotic solution $\tau_{\rm as}=[g, t_{\rm as}]$ to the almost maximal equation as presented in Equation~\eqref{am2as}, and again let $\m\stackrel{g}{:=}\nabla t_{\rm as}$ and
$\akappa\stackrel{g}{:=}-\tfrac14(\Delta+J)t_{\rm as}$,
 so that
\begin{equation}\label{anor}
 \m_a \m^a+2\akappa t_{\rm as}\stackrel{g}{=}-1+\mathcal O(t_{\rm as}^4),
\end{equation}
and 
\begin{equation} \label{acor}
\m_an^a+\akappa s+(\rho-c)t_{\rm as}\stackrel{g}{=}bs^3.
\end{equation}
We will study the behavior of the defining function $t_\mathrm{as}$ along $M$, so set
 $\tilde{t}_{\rm as}\stackrel{\tilde{g}}{:=}t_{\rm as}|_M$. Since $t_{\rm as}|_S=0$ and
$\m^a|_S=\hat{\bar{m}}^a$ it follows that 
$\tilde{t}_{\rm as}|_S=0$, $\tilde\nabla\tilde{t}_{\rm as}|_S=0$, $\nabla_{\hat m} t_\mathrm{as}\stackrel M = \hat m^a \m_a$, and
$\hat{\bar{m}}_a \m^a|_S=-1$.
Then 
\begin{equation}\label{m2mhat} \m^a
%=\nabla^a t_\mathrm{as} 
\stackrel{M}{=}\tilde\nabla^a\tilde{t}_{\rm as}-(\hat{m}^b \m_b)\hat{m}^a
\end{equation}
 and
\begin{equation}\label{pullbackhessian}
 \tilde\nabla_a\tilde\nabla_b\tilde{t}_{\rm as}
 \stackrel{M}{=}\tilde\top_a{}^c\tilde\top_b{}^d\nabla_c \m_d
   +(\hat{m}^c \m_c)\II^M_{ab}\, .
\end{equation}
Computing $\nabla_{\hat m}$ of 
%the normalization of $m$ in
  Equation~\eqref{anor} along $M$, and then using Equation~\eqref{m2mhat} as well as
 $\Delta t_{\rm as}\stackrel{g}{=}-4\akappa-Jt_{\rm as}$ and $g\stackrel M = \tilde g -\hat m\otimes \hat m$, gives
\[
 \tilde g^{ab}\nabla_a \m_b+3\akappa
 \stackrel{M}{=}-J\tilde{t}_{\rm as}+
 \frac{\tilde{t}_{\rm as}\,\hat{m}^a\nabla_a\akappa
       +\hat{m}^a(\nabla_a \m_b)\tilde\nabla^b\tilde{t}_{\rm as}}
      {\hat{m}^c \m_c}
 +\mathcal O_{\rm phg}(\tilde{t}_{\rm as}^3)\, ,
\]
where, as usual,  $\mathcal O_{\rm phg}$ refers to a polyhomogeneous expansion in $\tilde s$ and $\log \tilde s$. We now take the trace of Equation~\eqref{pullbackhessian}
and restrict Equation~\eqref{acor} to $M$.
Using $\hat m^a n_a\stackrel M=\tilde sH^M$
and rewriting the term $\akappa\tilde s$ using the
preceding display then gives
\begin{equation}\label{tildeteq}
 \left(\tilde n^a\tilde\nabla_a-\tfrac{\tilde s}{3}\tilde\Delta\right)\tilde{t}_{\rm as}
\stackrel{\tilde g}{=}b\tilde s^3-\tfrac{\tilde s\,\hat{m}^a(\nabla_a \m_b)\tilde\nabla^b\tilde{t}_{\rm as}}
       {3\hat{m}^c \m_c} + \mathcal O_{\rm phg}(\tilde{t}_{\rm as})\, .
\end{equation}
We can also determine the remainder term denoted by $\mathcal O_{\rm phg}(\tilde{t}_{\rm as})$ in
the above, and find that it equals
\[
 \left(c-\rho+\tfrac{\tilde s}{3}J
 -\tfrac{\tilde s\,\hat{m}^{a}\nabla_a\akappa}
        {3\hat{m}^{b} \m_b}\right)\tilde{t}_{\rm as}
 +\mathcal O_{\rm phg}(\tilde s\,\tilde{t}_{\rm as}^{3})\, .
\]
Since $\hat{m}^{a}\m_a\stackrel S=-1$, 
%all the coefficients in the Equation~\eqref{tildeteq} are bounded near $S$. 
it follows that $(\hat m^c \m_c)^{-1}$ is finite in a neighborhood of $S$.

We now use Equation~\eqref{tildeteq}
to determine the asymptotic behavior of  
$\tilde t_\mathrm{as}$.
The asymptotic solution~$\tilde{t}_{\rm as}$ need not vanish on $M$, so we would like to compute its leading behavior.  Hence assume this  has
 order ${\mathcal O}_\mathrm{phg}(\tilde s^k)$ for some $k \in \mathbb{N}$.
 Then 
the last two terms on right-hand side of Equation~\eqref{tildeteq}  are of order
$\mathcal O_{\rm phg}(\tilde s^k)$. To handle the left-hand side we use
$$
 \tilde\Delta\tilde s^k
 \stackrel{\tilde{g}}{=}k(k-1)\tilde s^{k-2}\tilde n^a\tilde n_a
  +k\tilde s^{k-1}\tilde\Delta\tilde s\, .
$$
and $\tilde n^a\tilde n_a\stackrel{\tilde{g}}{=}1+\mathcal O_{\rm phg}(\tilde s)$. It follows that
\begin{equation}\label{km4}
 \left(\tilde n^a\tilde\nabla_a-\tfrac{\tilde s}{3}\tilde\Delta\right)
 \tilde s^k 
% \stackrel{\tilde{g}}{=}\tfrac{k(4-k)}3\tilde s^{k-1}\tilde n^a\tilde n_a
%   -\tfrac{k}{3}\tilde s^k\tilde\Delta\tilde s 
\stackrel{\tilde{g}}{=}-\tfrac{k(k-4)}3\tilde s^{k-1}
   +\mathcal O_{\rm phg}(\tilde s^k)\, ,
\end{equation}
so, for $k\neq4$, a nonzero leading coefficient of
$\tilde t_{\rm as}$ at power $\tilde s^k$ produces a nonzero
term at order~$\tilde s^{k-1}$ on the left-hand side of
Equation~\eqref{tildeteq}.
Recalling that $\tilde t_{\rm as}|_S=0=
\tilde\nabla\tilde{t}_{\rm as}|_S$ we may assume that $k\geq 2$ (the analysis below also rules out the leading behavior $\tilde s^2 \log \tilde s$). Matching orders on the left and right sides of Equation~\eqref{tildeteq} shows that~$\tilde t_{\rm as}$ can have neither leading term of order $\tilde s^2$ nor  $\tilde s^3$. The vanishing of the indicial coefficient at $k=4$ means
that a smooth quartic term cannot match the order-$\tilde s^3$
obstruction term in Equation~\eqref{tildeteq}. The first
logarithmic correction therefore must be proportional to
$\tilde s^4\log\tilde s$, and is necessary wherever $b|_S\neq0$.
Observe that
$$
\left(\tilde n^a\tilde\nabla_a-\tfrac{\tilde s}{3}\tilde\Delta\right)
 (\tilde s^4\log\tilde s)\stackrel{\tilde{g}}{=}\tilde s^3\left(4\log\tilde s+1
                   -\tfrac{12\log\tilde s+7}{3}\right)
        +\mathcal O_{\rm phg}(\tilde s^4 \log \tilde s) \stackrel{\tilde{g}}{=}-\tfrac43\tilde s^3
        +\mathcal O_{\rm phg}(\tilde s^4 \log \tilde s)\, ,
$$
which fixes the coefficient of
$\tilde s^4\log\tilde s$ in $\tilde t_\mathrm{as}$ to be $-3b/4$.
Thus 
\begin{equation}\label{asymtas}
 \tilde{t}_{\rm as}\stackrel{\tilde{g}}{=}
 %\tilde %s^4\tilde{t}_{\rm as}^{(4)}
 -\tfrac34b\,\tilde s^4\log\tilde s +\mathcal O_{\rm phg}(\tilde s^4)\, .
\end{equation}
Equations~\eqref{tildeteq} and~\eqref{asymtas} together imply
\begin{equation} \label{tases}
 \left(
 \tilde n^a\tilde\nabla_a
 -\tfrac{\tilde s}{3}\tilde\Delta
 \right)
 \left(
 \tilde t_{\rm as}
 +\tfrac34b\,\tilde s^4\log\tilde s
 \right)
 \stackrel{\tilde g}{=}
 \mathcal O_{\rm phg}(\tilde s^4\log\tilde s)\, .
\end{equation}
The quantity in the second pair of parentheses is
$\mathcal O_{\rm phg}(\tilde s^4)$.
At order $\tilde s^4$, the right-hand side of
Equation~\eqref{tases}, as obtained from
Equation~\eqref{tildeteq}, is at most linear in $\log\tilde s$:
only smooth low-order coefficient jets acting on the known
quartic terms in Equation~\eqref{asymtas} contribute at this
order. Consequently, its order-four coefficient contains
no powers $(\log\tilde s)^p$ with $p\ge2$.
On the other hand, for $p\geq 2$,
\[
 \left(
 \tilde n^a\tilde\nabla_a
 -\tfrac{\tilde s}{3}\tilde\Delta
 \right)
 \big(\tilde s^5(\log\tilde s)^p\big)
 \stackrel{\tilde g}{=}
 -\tfrac53\tilde s^4(\log\tilde s)^p
 +\mathcal O_{\rm phg}
   \big(\tilde s^4(\log\tilde s)^{p-1}\big)\, .
\]
Thus, the quantity in the  second pair of parentheses on the left-hand side of
Equation~\eqref{tases} contains no terms
$\tilde s^5(\log\tilde s)^p$ with $p\geq2$. Hence, the remainder on the right-hand side of Equation~\eqref{asymtas} is ${\mathcal O}(\tilde s^4)+{\mathcal O}_\mathrm{phg}(\tilde s^5 \log \tilde s)$.

\smallskip

We  want to use the trace-free part of Equation~\eqref{pullbackhessian} to investigate the regularity of $\IIo^M\!$, and in particular identify the lowest order logarithms appearing in its boundary asymptotics. For that we need various ingredients:  
We first need to know $\hat m_a \m^a$ along $M$: The normal--tangential decomposition of $\m$, together with
$\hat m_a \hat m^a \stackrel{M}{=}-1$, gives
$$
 \m_a \m^a
 \stackrel{M}{=}
 |\tilde\nabla\tilde t_{\rm as}|_{\tilde g}^{\,2}
 -(\hat m^a \m_a)^2.
$$
Applying Equation~\eqref{anor} along $M$ then yields
\[
 (\hat m_a \m^a)^2
 \stackrel{M}{=}
 1+2\akappa\tilde t_{\rm as}
 +|\tilde\nabla\tilde t_{\rm as}|_{\tilde g}^{\,2}
 +\mathcal O_{\rm phg}(\tilde t_{\rm as}^{\,4})\, .
\]
Equation~\eqref{asymtas} implies that the right-hand side of the above display equals 
$1+\mathcal O_{\rm phg}(\tilde s^4 \log \tilde s )$.
Moreover, $\m^a|_S=\hat{\bar m}^{a}=\hat m^a|_S$, so
$\hat m_a \m^a\stackrel S=-1$. Hence
$$
 \hat m_a \m^a
 \stackrel{M}{=}
 -1+\mathcal O_{\rm phg}(\tilde s^4 \log \tilde s)\: \Rightarrow \:
 (\hat m_a \m^a)^{-1}
 \stackrel{M}{=}
 -1+\mathcal O_{\rm phg}(\tilde s^4 \log \tilde s)\, .
$$
We now turn to the tensor $\nabla_a \m_b$ appearing in Equation~\eqref{pullbackhessian}. 
Since it is smooth on $X$, its pullback to~$M$ first produces logarithms at order ${\mathcal O}_\mathrm{phg}(\tilde s^3\log \tilde s)$ so
the term $\tilde\top_a{}^c\tilde\top_b{}^d\nabla_c \m_d$ in Equation~\eqref{pullbackhessian}
has no contribution of order ${\mathcal O}_\mathrm{phg}(\tilde s^2\log \tilde s)$. We may now turn to  the Hessian of $\tilde t_{\rm as}$ along $M$ which---see Equation~\eqref{asymtas}---requires that we  compute two derivatives of $\tilde s^4\log\tilde s$.
Recalling that  $\tilde n_a=\tilde\nabla_a\tilde s$, we find
$$
 \tilde\nabla_a\tilde\nabla_b(\tilde s^4\log\tilde s) \stackrel{\tilde g}{=}
 \tilde s^2(12\log\tilde s+7)\tilde n_a\tilde n_b +\tilde s^3(4\log\tilde s+1)\tilde\nabla_a\tilde n_b\, .
$$
Consequently, 
$$
 \tilde\nabla_a\tilde\nabla_b\tilde t_{\rm as}
 \stackrel{\tilde g}{=}
 -9b \tilde n_a\tilde n_b  \,\tilde s^2\log\tilde s\,  +\mathcal O_{\rm phg}(\tilde s^2)\, .
$$
The trace-free part of
$\tilde n_a\tilde n_b$ is $ |\tilde n|_{\tilde g}^2
 \big(
 \tfrac{\tilde n_a\tilde n_b}{|\tilde n|_{\tilde g}^2}
 -\tfrac13\tilde g_{ab}
 \big)$
and
 $|\tilde n|_{\tilde g}^2
=1+\mathcal O_{\rm phg}(\tilde s)$.
It follows that 
$$
 \tfrac{\tilde\nabla_{(a}\tilde\nabla_{b)^{\circ}}\tilde t_{\rm as}}
      {\hat m_c \m^c}=
      9b  \tilde s^2\log\tilde s\:
 \bigg(
 \tfrac{\tilde n_a\tilde n_b}{|\tilde n|_{\tilde g}^2}
 -\tfrac13\tilde g_{ab}
 \bigg)
 +\mathcal O_{\rm phg}(\tilde s^2)
\, .
$$
Orchestrating the above we have that
\begin{equation}\label{klog}
 \IIo^M_{ab}
 \stackrel{\tilde g}{=}\IIo^{M,\mathrm{as}}_{ab}+\tilde s^2A_{ab}+9b  \tilde s^2\log\tilde s\:
 \bigg(
 \tfrac{\tilde n_a\tilde n_b}{|\tilde n|_{\tilde g}^2}
 -\tfrac13\tilde g_{ab}
 \bigg)
 +\mathcal O_{\rm phg}(\tilde s^3 \log \tilde s),
\end{equation}
where
$\IIo^{M,\mathrm{as}}$ 
solves Equation~\eqref{aresidual},
and $A$ is a smooth tensor not determined by solving 
that equation asymptotically. We can now insert the expansion in Equation~\eqref{klog}  into the  almost momentum constraint
and compare the result with Equation~\eqref{aresidual} to obtain
$$
 0\stackrel{\tilde g}{=}
 \tilde s^2\big(L_a+6b \tilde n_a\big)
 +\mathcal O_{\rm phg}(\tilde s^3 \log \tilde s)\, .
$$
Remembering that
$\tilde n_a|_S=\hat{\tilde n}_a$, proves that $\bar L_a\stackrel{\tilde g}{=}-6b|_S\hat{\tilde n}_a$ as claimed.

\section{Proof of Theorem \ref{mainth}}\label{apenmain}

Here we show how to re-express $\IVo^{\scri}_{\hat{\bar m}a}$ in terms of the geometry of $M \hookrightarrow X$ along~$S$ and thereby establish Equation~\eqref{4toh4p2} required for the proof of Theorem~\ref{mainth} given in Section~\ref{proofth1}.
To start with
we fix a smooth ambient metric $g\in\bm c$ and let $\tilde g$ be its
pullback to the maximal hypersurface $M$. We will also use the notation
of Appendices~\ref{apenh4}--\ref{apenwillmore} and the polyhomogeneous boundary expansions described in
Section~\ref{notreg}.
The case where the embedding $M\hookrightarrow X$ is smooth follows directly from
Appendices~\ref{apenh4} and~\ref{apenp2ren}. Hence we need to show that all requisite 
identities hold when logarithms   are present in the asymptotics of the defining density for $M\hookrightarrow X$ and hence also $\tilde g$ and $\IIo^M$. These logarithms were already partly analyzed in Section~\ref{notreg} and Appendix~\ref{apenwillmore}.

First consider $h_4^{\rm ren}$, which probes the jets of $\tilde s$ along $S$ in directions normal to $S\hookrightarrow (M,\tilde g)$. 
Let us therefore  compute the Hessian of  $\tilde s$. Because $\tilde s$ is the pullback of $s$ we  use the 
almost Einstein Equation~\eqref{aeeq}  for the embedding $M\hookrightarrow (X,\bm c)$ and find
\begin{equation}\label{apend-hessian}
 \tilde\nabla_a\tilde\nabla_b\tilde s
 \stackrel{M}{=}
 -\tilde s\,\tilde\top_a{}^c\tilde\top_b{}^dP_{cd}
 -\rho\,\tilde g_{ab}
 +\tilde s H^M\II^M_{ab}\, .
\end{equation}
Here we used $\hat m_a n^a\stackrel M=\tilde sH^M$ from the almost maximal Equation~\eqref{am2}.
Note that the second fundamental form
$\II^M$ has its first possible
logarithm at order~$\tilde s^2\log\tilde s$, since it involves
at most two derivatives of the embedding (recall that the leading log behavior of the function $\tilde t_\mathrm{as}$ is $\tilde s^4 \log \tilde s$; see the discussion in Appendix~\ref{apenwillmore}).

Next we differentiate Equation~\eqref{apend-hessian} twice 
with  $\tilde \nabla$ along $M$.
The pullback of the ambient Schouten~$P$ to $M$ has continuous second derivatives along $M$, as does the
restriction of $\rho$ (recall that the first logarithms in $\tilde g$ appear at order $\tilde s^4 \log \tilde s$). The last term on the right-hand side of Equation~\eqref{apend-hessian} 
has leading logarithms at order $\tilde s^3 \log\tilde s$ because $\II^M$ has leading
logarithms at order~$\tilde s^2\log\tilde s$.
It follows that $\tilde \nabla_a \tilde \nabla_b \tilde \nabla_c \tilde \nabla_d \tilde s$ is continuous along $S$.

The boundary calculations in Appendix~\ref{apenh4} use the first
boundary jet of $\II^M$, the fourth derivative of $\tilde s$ just considered, as well as  identities relying on the maximality
condition. These required at most three derivatives of the defining function $t$. It follows that all the computations involving~$h_4^\mathrm{ren}$ given in Appendix~\ref{apenh4} leading first to Equation~\eqref{h4renappen} and in turn its relation  to Equation~\eqref{h4renA},
hold without modification. Equation~\eqref{h4renA} is required in part to obtain the  Ashtekar--Das expression for the charge $Q_{\bar \xi}$, while Equation~\eqref{h4renappen} partly gives its Cauchy hypersurface expression.

\medskip

It remains to show that $p_2^{\rm ren}$ is also unaffected by logarithms. For that we start with the asymptotic formula for  $\IIo^M$ given in Equation~\eqref{klog}.
The terms $\IIo^{M,\mathrm{as}}+\tilde s^2A$
on the right-hand side are smooth. The term $A$---undetermined 
when solving only for an asymptotic solution---captures  non-local data required for constructing charges and in particular $p_2^\mathrm{ren}$. Our current task is to show that the  log term in Equation~\eqref{klog} does not change $p_2^\mathrm{ren}$. For that, we introduce the extension $\check n$ of  $\hat {\tilde n}|_S$
to a collar neighborhood of $S$ in  $M$ given by the  unit normal to  level sets of
$\tilde s$, and extend the surface projector and inverse surface
metric accordingly:
\[
 \check n^a\stackrel{\tilde g}{:=}
       \tfrac{\tilde n^a}{|\tilde n|_{\tilde g}},\qquad
 \check\top_a{}^b
       :=\delta_a{}^b-\check n_a\check n^b,\qquad
 \check g^{ab}
       :=\tilde g^{ab}-\check n^a\check n^b.
\]
%These agree with the original boundary fields at $S$. 
Defining also the trace-free part of $\check n \otimes \check n$ by
$
 Q_{ab}\stackrel{\tilde g}{:=}
        \check n_a  \check n_b       -\tfrac13\tilde g_{ab}
%       \tfrac{\tilde n_a\tilde n_b}{|\tilde n|^2_{\tilde g}}
%      -\tfrac13\tilde g_{ab}
$, we have that
\begin{equation}\label{apend-logprojection}
 \tilde n^bQ_{ab}\stackrel{\tilde g}{=}\tfrac23\tilde n_a\, ,
 \qquad
 \check\top_a{}^e\check n^dQ_{de}=0\ , 
\end{equation}
throughout the collar.

We need to show that the leading transverse order term
$\tilde{\bar\top}_a{}^e
 \hat{\tilde n}^b\hat{\tilde n}^c\hat{\tilde n}^d
 \tilde\nabla_b\tilde\nabla_c\IIo^M_{de}|_S$
in $p_2^\mathrm{ren}$
 is well defined. Therefore 
  we now compute
\[
 \tilde\nabla_b\tilde\nabla_c
       \big(b \tilde s^2\log\tilde s\,Q_{de}\big)
 \stackrel{\tilde g}{=}
 2b \big(\log\tilde s+\tfrac32\big)\tilde n_b\tilde n_cQ_{de}
 +\mathcal O_{\rm phg}(\tilde s \log \tilde s)\, .
\]
Upon contracting with $\check n^b\check n^c\check n^d$, projecting
 the remaining index to directions tangent to  the level sets of $\tilde s$, 
 and using Equation~\eqref{apend-logprojection}, the first term on the right-hand side of the above display vanishes. 
 Well definedness now follows. In fact, by inspection, we now see that $p_2^\mathrm{ren}$ as given in Equation~\eqref{p2ren} is also well defined.
 
 It only remains to verify that the tensor $p_2^\mathrm{ren}$ is still given by Equation~\eqref{4toh4p2}. For that we need only establish Equation~\eqref{ivp2ren}. In the absence of logarithms, this was already achieved by differentiating the
momentum constraint~\eqref{almcon} twice
in directions normal to $S$, and then evaluating the resulting tensor  along $S$ in directions tangent to $S$. The main identities required were given in Appendix~\ref{apenp2ren}.
In fact the derivation given there still holds in the presence of logs:
 That  $\,\tilde{\bar\top}_a{}^e
\tilde s\hh {\tilde n}^b{\tilde n}^c
\tilde\nabla_b\tilde\nabla_c\tilde\nabla^d\IIo^M_{de}$ is 
 well defined along~$S$ follows from a computation very similar to that we used to handle $\tilde{\bar\top}_a{}^e
 \hat{\tilde n}^b\hat{\tilde n}^c\hat{\tilde n}^d
 \tilde\nabla_b\tilde\nabla_c\IIo^M_{de}\big|_S$. The same applies to two derivatives of the second term $\tilde n^a\IIo^{M}_{ab}|_S$ in the almost momentum constraint equation.
 It remains to show that the identities in Equations~\eqref{p2firstjets},~\eqref{p2cotton} and~\eqref{p2mixedderiv} are still valid. They involve up to two derivatives of the almost momentum constraint, so this can easily be achieved using the same methods as above.


\begin{thebibliography}{10}

\bibitem{Alex}
S. Alexakis and R. Mazzeo, \emph{Renormalized area and properly embedded minimal surfaces in hyperbolic 3-manifolds}, Comm. Math. Phys. 297(3), 621--651 (2010).

\bibitem{chrand}
L. Andersson and P. T. Chru\'sciel,
\emph{Solutions of the constraint equations in general relativity satisfying
``hyperboloidal boundary conditions''},
Dissertationes Math. (Rozprawy Mat.) \textbf{355}, 1--100, 1996.

\bibitem{ACF}
L. Andersson, P. T. Chru\'sciel, and H. Friedrich,
\emph{On the regularity of solutions to the Yamabe equation and the existence
of smooth hyperboloidal initial data for Einstein's field equations},
Commun. Math. Phys. \textbf{149}(3), 587--612, 1992.

\bibitem{and}
M. T. Anderson,
\emph{$L^2$ curvature and volume renormalization of AHE metrics on
4-manifolds},
Math. Res. Lett. \textbf{8}(1--2), 171--188, 2001.

\bibitem{AvMcO}
P. Aviles and R. C. McOwen,
\emph{Conformal deformation to constant negative scalar curvature on
noncompact Riemannian manifolds},
J. Differential Geom. \textbf{27}(2), 225--239, 1988.

\bibitem{gwa}
C. Arias, A. R. Gover, and A. Waldron,
\emph{Conformal geometry of embedded manifolds with boundary from universal
holographic formulae},
Adv. Math. \textbf{384}, 107700, 2021.

\bibitem{abk}
A. Ashtekar, B. Bonga, and A. Kesavan,
\emph{Asymptotics with a positive cosmological constant: I. Basic framework},
Class. Quantum Grav. \textbf{32}(2), 025004, 2015.

\bibitem{adas}
A. Ashtekar and S. Das,
\emph{Asymptotically Anti-de Sitter space-times: Conserved quantities},
Class. Quantum Grav. \textbf{17}(2), L17--L30, 2000.

\bibitem{bfs}
M. Bianchi, D. Z. Freedman, and K. Skenderis,
\emph{Holographic renormalization},
Nucl. Phys. B \textbf{631}(1--2), 159--194, 2002.

\bibitem{CFFs}
S. Blitz, A. R. Gover, and A. Waldron,
\emph{Conformal fundamental forms and the asymptotically
Poincar\'e--Einstein condition},
Indiana Univ. Math. J. \textbf{72}(6), 2215--2284, 2023.

\bibitem{d2n}
S. Blitz, A. R. Gover, J. Kopi\'nski, and A. Waldron,
\emph{The Dirichlet-to-Neumann map for Poincar\'e--Einstein fillings},
Rev. Mat. Iberoam. \textbf{42}(1), 229--260, 2026.

\bibitem{bvv}
J. de Boer, E. P. Verlinde, and H. L. Verlinde,
\emph{On the holographic renormalization group},
JHEP \textbf{08}, 003, 2000.

\bibitem{cg}
A. \v{C}ap and A. R. Gover, \emph{A boundary-local mass cocycle and the mass of asymptotically hyperbolic manifolds}, Comm. Math. Phys. \textbf{405} 10, 233, 2024.

\bibitem{Cederbaum}
C. Cederbaum, J. Cortier and A. Sakovich, \emph{On the center of mass of asymptotically hyperbolic initial data sets}, Ann. Henri Poincar\'e \textbf{17}(6), 1505--1528, 2016.

\bibitem{Cherrier}
P. Cherrier, \emph{Probl\`emes de Neumann non lin\'eaires sur les vari\'et\'es riemanniennes}, J. Funct. Anal. \textbf{57} 2, 154--206, 1984.

\bibitem{chrh}
P. T. Chru\'sciel and M. Herzlich,
\emph{The mass of asymptotically hyperbolic Riemannian manifolds},
Pacific J. Math. \textbf{212}(2), 231--264, 2003.

\bibitem{chrnag}
P. T. Chru\'sciel and G. Nagy,
\emph{The mass of spacelike hypersurfaces in asymptotically anti-de Sitter
space-times},
Adv. Theor. Math. Phys. \textbf{5}(4), 697--754, 2001.

\bibitem{chrmt}
P. T. Chru\'sciel, D. Maerten, and P. Tod,
\emph{Rigid upper bounds for the angular momentum and centre of mass of
non-singular asymptotically anti-de Sitter space-times},
JHEP \textbf{11}, 084, 2006.

\bibitem{chrwut}
P. T. Chru\'sciel and R. Wutte,
\emph{Holographic is Hamiltonian, relatively},
Lett. Math. Phys. \textbf{116}, 109, 2026.

\bibitem{chrgal}
P. T. Chru\'sciel and G. J. Galloway,
\emph{Maximal hypersurfaces in asymptotically Anti-de Sitter spacetimes},
Celebratio Mathematica, 2023; arXiv:2208.09893 [gr-qc].

\bibitem{govk}
A. R. Gover and J. Kopi\'nski,
\emph{Higher fundamental forms of the conformal boundary of asymptotically
de Sitter spacetimes},
Class. Quantum Grav. \textbf{40}, 015001, 2023.

\bibitem{Goal}
A. R. Gover, \emph{Almost Einstein and Poincar\'e--Einstein manifolds in Riemannian signature}, J. Geom. Phys. \textbf{60}(2), 182--204, 2010.

\bibitem{bcs}
A. R. Gover, J. Kopi\'nski, and A. Waldron,
\emph{Boundary Curvature Scalars on Conformally Compact Manifolds},
J. Phys. A: Math. Theor. \textbf{58}(31), 315402, 2025.

\bibitem{gwln}
A. R. Gover and A. Waldron,
\emph{Conformal hypersurface geometry via a boundary
Loewner--Nirenberg--Yamabe problem},
Commun. Anal. Geom. \textbf{29}(4), 779--836, 2021.

\bibitem{will}
A. R. Gover and A. Waldron,
\emph{A calculus for conformal hypersurfaces and new higher Willmore energy
functionals},
Adv. Geom. \textbf{20}(1), 29--60, 2020.

\bibitem{gra1}
C. R. Graham,
\emph{Volume and area renormalizations for conformally compact Einstein
metrics},
in Proceedings of the 19th Winter School ``Geometry and Physics''
(Srn\'i, 1999), Rend. Circ. Mat. Palermo (2) Suppl. No. \textbf{63},
31--42, 2000.

\bibitem{gra2}
C. R. Graham,
\emph{Dirichlet-to-Neumann map for Poincar\'e--Einstein metrics},
Oberwolfach Rep. \textbf{2}(3), 2200--2203, 2005.

\bibitem{grar}
C. R. Graham and N. Reichert,
\emph{Higher-dimensional Willmore energies via minimal submanifold
asymptotics},
Asian J. Math. \textbf{24}(4), 571--610, 2020.

\bibitem{grahwit}
C. R. Graham and E. Witten,
\emph{Conformal anomaly of submanifold observables in AdS/CFT
correspondence},
Nucl. Phys. B \textbf{546}(1--2), 52--64, 1999.

\bibitem{hss}
S. de Haro, K. Skenderis, and S. N. Solodukhin,
\emph{Holographic reconstruction of spacetime and renormalization in the
AdS/CFT correspondence},
Commun. Math. Phys. \textbf{217}(3), 595--622, 2001.

\bibitem{HenSk}
M. Henningson and K. Skenderis,
\emph{The holographic Weyl anomaly},
JHEP \textbf{07}, 023, 1998.

\bibitem{him}
S. Hollands, A. Ishibashi, and D. Marolf,
\emph{Comparison between various notions of conserved charges in
asymptotically AdS spacetimes},
Class. Quantum Grav. \textbf{22}(14), 2881--2920, 2005.

\bibitem{iw}
V. Iyer and R. M. Wald,
\emph{Some properties of the Noether charge and a proposal for dynamical black
hole entropy},
Phys. Rev. D \textbf{50}(2), 846--864, 1994.

\bibitem{khurikopinski}
M. Khuri and J. Kopi\'nski,
\emph{Asymptotically hyperbolic Einstein constraint equations with
apparent horizon boundary and the Penrose inequality for perturbations
of Schwarzschild-AdS},
Class. Quantum Grav. \textbf{40}(4), 045007, 2023.

\bibitem{ln}
C. Loewner and L. Nirenberg,
\emph{Partial differential equations invariant under conformal or projective
transformations},
in \emph{Contributions to Analysis: A Collection of Papers Dedicated to
Lipman Bers}, Academic Press, New York--London, 245--272, 1974.

\bibitem{Mazz}
R. Mazzeo. \emph{Regularity for the singular Yamabe problem}, Indiana Univ. Math. J., \textbf{40}(4):1277--1299, 1991.

\bibitem{michel}
B. Michel, \emph{Geometric invariance of mass-like asymptotic invariants}, J. Math. Phys. \textbf{52} 5, 052504. 2011.

\bibitem{penr}
R. Penrose and W. Rindler,
\emph{Spinors and Space-Time, Vol. 1: Two-Spinor Calculus and Relativistic
Fields},
Cambridge Monographs on Mathematical Physics, Cambridge University Press,
Cambridge, 1984.

\bibitem{Polyakov}
A. Polyakov, \emph{Fine structure of strings}, Nuclear Phys. B, \textbf{268}(2), 406--412, 1986.

\bibitem{kl}
K. Skenderis,
\emph{Lecture notes on holographic renormalization},
Class. Quantum Grav. \textbf{19}(22), 5849--5876, 2002.

\bibitem{wz}
R. M. Wald and A. Zoupas,
\emph{General definition of ``conserved quantities'' in general relativity
and other theories of gravity},
Phys. Rev. D \textbf{61}(8), 084027, 2000.

\bibitem{Willmore}
T.~J. Willmore, \emph{Note on embedded surfaces},
An. \c{S}tiin\c{t}. Univ. ``Al. I. Cuza'' Ia\c{s}i Sect. I a Mat.
(N.S.) \textbf{11B}, 493--496, 1965.

\bibitem{wang}
X. Wang,
\emph{The mass of asymptotically hyperbolic manifolds},
J. Differential Geom. \textbf{57}(2), 273--299, 2001.

\bibitem{york}
J. W. York, Jr.,
\emph{Conformally invariant orthogonal decomposition of symmetric tensors on
Riemannian manifolds and the initial-value problem of general relativity},
J. Math. Phys. \textbf{14}(4), 456--464, 1973.

\end{thebibliography}
\end{document}